\documentclass[amsmath,amssymb,amsfonts,aps,physrev,preprint,superscriptaddress]{revtex4-1}
\usepackage[utf8]{inputenc}
\usepackage[british]{babel}
\usepackage{graphicx}
\usepackage{bm}
\usepackage{amsmath}
\usepackage{amssymb}
\usepackage{subcaption}
\usepackage{xcolor}
\usepackage{hyperref}
\usepackage{marginnote}
\usepackage{siunitx}

\newcommand{\bx}{\bm{x}}
\newcommand{\bp}{\bm{p}}
\newcommand{\bn}{\bm{\nabla}}

\newcommand{\bd}{\bm{d}}
\renewcommand{\bf}{\bm{f}}
\newcommand{\bu}{\bm{u}}
\newcommand{\bU}{\bm{U}}
\newcommand{\bO}{\bm{\Omega}}
\newcommand{\bF}{\bm{F}}
\newcommand{\bG}{\bm{G}}
\newcommand{\Oseen}{\bm{\mathcal{G}}}

\begin{document}
	
	\title{Direct vs indirect hydrodynamic interactions during  bundle formation of bacterial flagella}
	\date{\today}
	\author{Alexander Chamolly}
	\email{alexander.chamolly@ens.fr}
	\affiliation{Department of Applied Mathematics and Theoretical Physics, University of Cambridge, Wilberforce Road, Cambridge CB3 0WA, UK}
	\affiliation{Laboratoire de Physique de l'Ecole Normale sup\'erieure, 24 Rue Lhomond, 75005 Paris, France}
	\author{Eric Lauga}
	\email{e.lauga@damtp.cam.ac.uk}
	\affiliation{Department of Applied Mathematics and Theoretical Physics, University of Cambridge, Wilberforce Road, Cambridge CB3 0WA, UK}
	
	\begin{abstract}
	
Most motile bacteria swim in viscous fluids by rotating multiple helical flagellar filaments. These semi-rigid filaments repeatedly join (`bundle') and separate (`unbundle'), resulting in a two-gait random walk-like motion of the cell. In this process, hydrodynamic interactions between the filaments are  known to play an important role and can be categorised into two distinct types: {\it direct} interactions mediated through flows that are generated through the actuation of the filaments themselves, and  {\it indirect} interactions mediated through the motion of the cell body (i.e.~flows induced in the swimming frame that result from propulsion).
To understand the relative importance of these two types of interactions, we  study a minimal singularity model of flagellar bundling. Using hydrodynamic images, we solve for the flow analytically and compute both direct and indirect interactions exactly as a function of the length of the flagellar filaments and their angular separation. We show  (i) that the generation of thrust by flagella alone is sufficient to drive the system towards a bundled state through both types of interaction in the entire geometric parameter range; (ii) that for both thrust- and rotation-induced flows indirect advection dominates for long filaments and at wide separation, i.e.~primarily during the early stages of the bundling process; and (iii) that, in contrast, direct interactions dominate when flagellar filaments are in each other's wake, which we characterise mathematically. We further introduce a numerical elastohydrodynamic model that allows us to compute the dynamics of the helical  axes of each flagellar filament while analysing direct and indirect interactions separately. With this we show (iv) that the shift in balance between direct and indirect interactions is non-monotonic during the bundling process, with a peak in direct dominance, and that different sections of the flagella are affected by these changes to different extents.

 	\end{abstract}

	\maketitle		
		
	\section{Introduction}
	
	Bacteria are present on Earth in great abundance in all kinds of environments, including on land~\cite{gans2005computational,gorbushina2009microbiology}, in the ocean~\cite{azam1983ecological,staley1999poles} and inside living hosts~\cite{ottemann1997roles}. Their study is not only important in order to understand many pathogenic diseases, but also serves as a model system for the locomotion of microorganisms and active particles in fluids~\cite{berg2008coli,lauga2009hydrodynamics,lauga2016bacterial} and their responses to environmental cues~\cite{berg1975chemotaxis,ottemann1997roles}. Moreover, their collective dynamics provide  one of the smallest model systems for the physics of active matter~\cite{ramaswamy2010mechanics,marchetti2013hydrodynamics}.
 	
	The majority of swimming bacteria are powered by so-called flagella.  Many bacteria such as the model organism \emph{Escherichia coli} (\emph{E.~coli}) are peritrichous, meaning that they  are driven by several flagella. These are  distributed almost randomly over the cell body, with a  slight bias due to the history of cell division~\cite{guttenplan2013cell}. Each flagellum consists of a long passive filament connected  at its base to a specialised rotary molecular motor through a short flexible hook~\cite{turner2000real,berg2008coli,berg1973bacteria}. The flagellar filaments are slender helices, assembled from proteins that form a relatively rigid structure, while the hook is significantly more flexible and acts both as a mechanical link and as a torsional spring. 
	
\begin{figure}[t]
		\centering
		\includegraphics[width=0.6\textwidth]{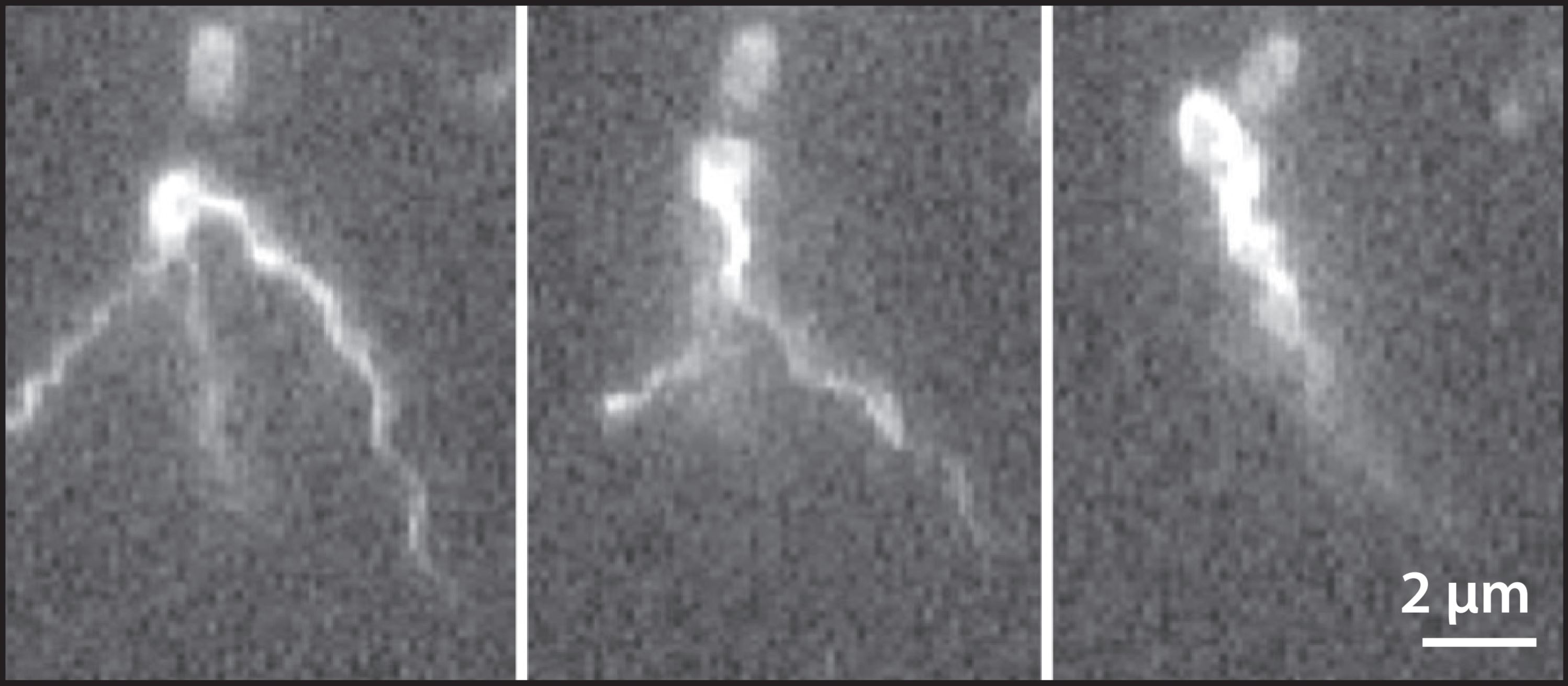}
		\caption{Snapshots of the bundling process occurring during the motion of \emph{E.~coli} bacteria wherein counter-clockwise rotating helical flagellar filaments arrange to form a bundle behind the cell body. Adapted from  Turner,  Ryu and   Berg  (2000)   ``Real-Time Imaging of Fluorescent Flagellar Filaments''
		{\it J. Bacteriol.}, $\bm{182}$, 2793-2801, with permission~\cite{turner2000real}.
 }\label{bund:fig:darnton}
	\end{figure}

	Although flagellar motors are actuated independently with no known biological coordination, it is observed experimentally that   bacteria are able to perform a two-gait ``run-and-tumble'' motion to explore their  fluid environment~\cite{berg1993random}. During a run event, the flagellar filaments co-rotate in a counter-clockwise (CCW) fashion and form a coherent helical bundle behind the cell, pushing it forward and leading to swimming. During a tumble event at least one of the motors switches to rotate clockwise (CW), leading to unbundling, i.e.~dispersion of the filament assembly, and a random reorientation of the cell. After a short period, the reversed motor switches back to CCW rotation, the bundle reforms and the process repeats~\cite{turner2000real,darnton2007torque}. An illustration of the bundling process in \emph{E.~coli} experiments  is shown in Fig.~\ref{bund:fig:darnton}~\cite{turner2000real}.
	
	Unlike the flagella of eukaryotic organisms such as the green alga \emph{Chlamydomonas reinhardtii}~\cite{polin2009chlamydomonas}, ciliates such as \emph{Paramecium}~\cite{braybook}, and spermatozoa~\cite{ishijima1986flagellar}, which are able to generate internal forces through the contraction of molecular motors within their flagella~\cite{lindemann1994geometric,lindemann1994model,hilfinger2009nonlinear,braybook}, the flagellar filaments of bacteria are passive, and the rotation of the flagella motor is the exclusive source of dynamic forcing~\cite{berg1973bacteria,silverman1974flagellar}. Nevertheless, the  dynamics of flagellated bacteria are a highly complex elastohydrodynamic problem~\cite{lauga2016bacterial}. The protein structure of the filaments allows for multiple different helical configurations, a phenomenon called polymorphism~\cite{calladine1978change,srigiriraju2006model}, which may change depending on the dynamic load~\cite{darnton2007force}. Moreover, because the flagellar filaments and especially the hook are flexible, and the interaction between these slender elastic structures, the cell body and the fluid lead to non-linearities that are difficult to compute~\cite{katsamba2019propulsion}.
	
	Despite these challenges, the dynamics of bundling  and unbundling have  been the subject of numerous theoretical~\cite{kim2004hydrodynamic,man2016hydrodynamic,riley2018swimming,tatulea2020geometrical}, numerical \cite{reigh2012synchronization,gebremichael2006mesoscopic,janssen2011coexistence,flores2005study,adhyapak2015zipping,eisenstecken2016bacterial,ishimoto2019n,lim2012fluid,nguyen2018impacts,lee2018bacterial,watari2010hydrodynamics,reichert2005synchronization} and experimental~\cite{berg2003rotary,berg2008coli,kim2003macroscopic,kim2004particle,macnab1977bacterial,qu2018changes,darnton2007torque,turner2010visualization,turner2000real} studies. Early studies examined the dynamics of filaments without taking into account the cell body~\cite{reigh2012synchronization,gebremichael2006mesoscopic,kim2003macroscopic,kim2004hydrodynamic,flores2005study,janssen2011coexistence,man2016hydrodynamic,lim2012fluid}, or ignored elastic effects~\cite{reichert2005synchronization}. With advances in computational power and the sophistication of numerical methods~\cite{cortez2001method,cortez2018regularized,hall2019efficient}, more recent work has been able to accurately simulate the swimming dynamics of multi-flagellated motile bacteria~\cite{adhyapak2015zipping,ishimoto2019n,nguyen2018impacts}. Some studies also focus on a particular aspect of the problem, such as flagellar polymorphism~\cite{lee2018bacterial} or the geometrical constraints of entanglement~\cite{tatulea2020geometrical}. 
		
	In this paper we address   the early stages of bundle formation, and aim to understand  fundamentally which hydrodynamic effects   dominate in the process  of flagellar filament assembly. Several different possible candidates have been proposed in the  literature. For instance, it was shown numerically and experimentally that the co-rotation of two elastic helices orthogonal to an interface leads to wrapping and mutual attraction~\cite{kim2003macroscopic,kim2004hydrodynamic}, while the wrapping occurs also for rotating straight filaments in an unbounded fluid~\cite{man2016hydrodynamic}. In both cases the effect is due to {\it direct} hydrodynamic interactions between the filaments.
	
	Other models have focused on the role of the cell body. It was  found   theoretically~\cite{powers2002role} and numerically~\cite{adhyapak2015zipping} that the motility of the cell body enables bundle formation through a ``zipping'' effect. The existence of an elastohydrodynamic instability due to a balance between fluid drag and  hook elasticity has also been demonstrated~\cite{riley2018swimming}. The crucial difference between these examples and those above is that bundling here does not occur as a result of  {\it direct } hydrodynamic interactions between the flagellar filaments, but rather {\it indirectly} through flows that are generated in the swimming frame due to  rotation and translation of the cell body.
	 	
	With this physical description in mind, we   introduce a minimal hydrodynamic model of flagellar bundling in \S\ref{bund:sec:model}, which  allows us to make precise the notion of direct vs indirect interactions.	We then proceed to analyse the two  hydrodynamic contributions in details, first for the flows due to flagellar thrust in \S\ref{bund:sec:thrust}, and then for the flows due to their rotation in \S\ref{bund:sec:rot}.  We next introduce a numerical elastohydrodynamic rod-and-spring model for the axis  of flagellar filaments in \S\ref{bund:sec:numerics} that allows us to analyse the dominant hydrodynamic contributions to the early bundling process for a model bacterium in detail. 	We conclude with a discussion of our results in \S \ref{bund:sec:discussion}.

	\begin{figure}[t]
		\centering
		\includegraphics[width=0.4\textwidth]{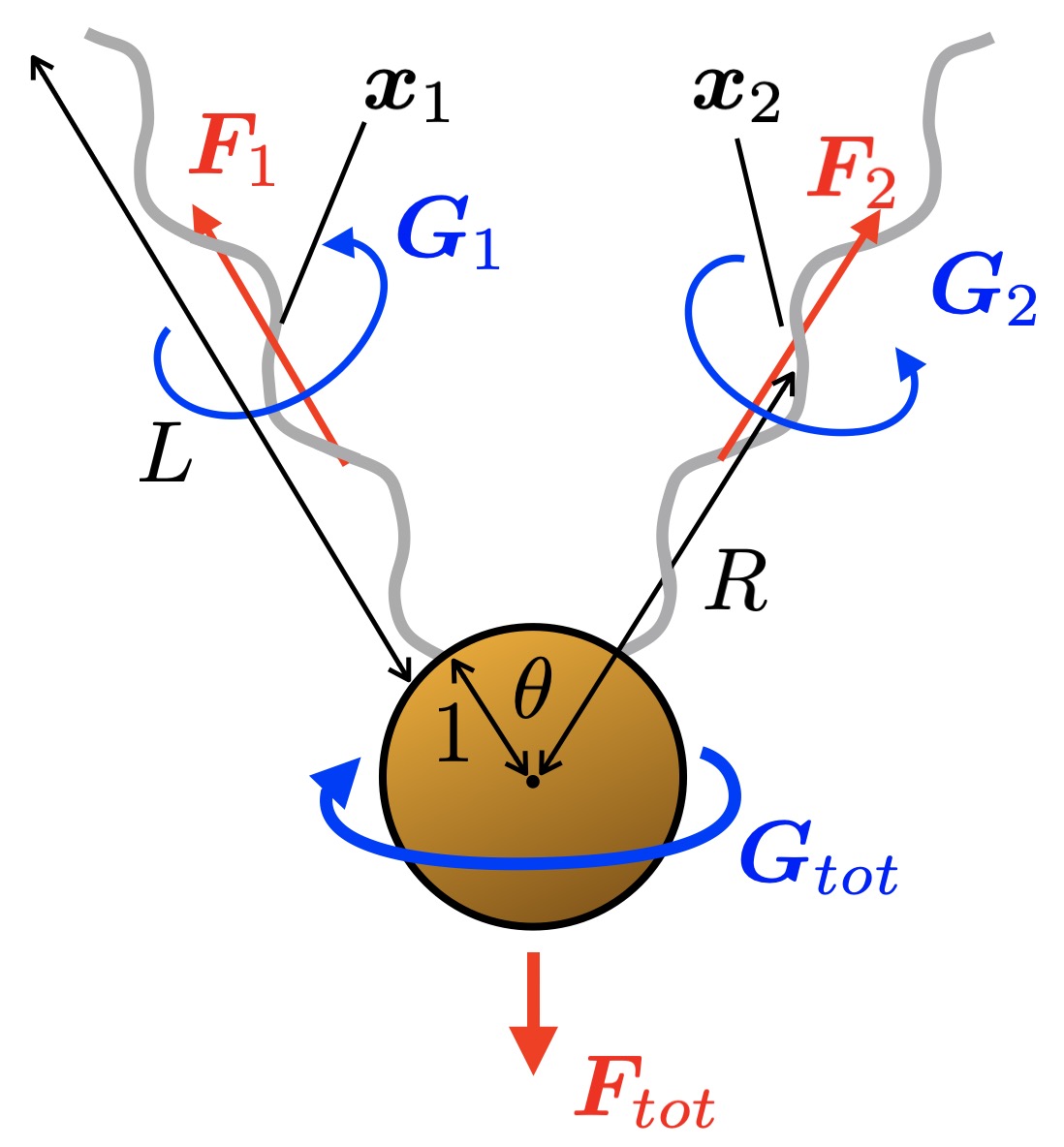}
		\caption{Minimal hydrodynamic model for flagellar bundling. The bacterium is modelled as a rigid sphere of unit radius, and equipped with two flagella. The flagellar filaments have axial length $L$ and are separated by a solid angle $\theta$ at their base. Labelling the filaments by $i=\{1,2\}$, each is modelled as a radial point force $\bF_i$ and point torque $\bG_i$ acting on the surrounding fluid and placed at $\bx_i$, which is at a distance $R=L/2+1$ from the centre of the cell body at the coordinate origin. Since the swimming cell is force- and torque-free, the cell body exerts a counterforce and -torque on the fluid.}
		\label{bund:fig:setup}
	\end{figure}

	\section{Minimal model of flagellar bundling}\label{bund:sec:model}
	
	In this section we introduce  a minimal model for the advection of bacterial flagellar filaments by the flow fields created by their actuation and the swimming of the cell. While idealised in its structure, the simplicity of the model allows us to study its dynamics analytically and to gain a quantitative understanding of the hydrodynamic forces at play.

	\subsection{Hydrodynamics of flagellar propulsion}

	Due to the microscopic size of bacteria, the fluid dynamics of their locomotion in a Newtonian viscous fluid are accurately described by the incompressible Stokes equations~\cite{purcell1977life,lauga2016bacterial},
	\begin{align}\label{bund:eq:Stokes}
	\bm{0}=-\bn p+\mu\nabla^2\bu+\bm{f},\quad \bn\cdot\bu =0,
	\end{align}
	where $\bu(\bx)$ is the flow velocity field, $p(\bx)$ the dynamic pressure, $\bm{f}(\bx)$ a body force density and $\mu$ the dynamic viscosity. Since the Stokes equations are linear, they allow for solution approaches using Green's functions (termed Stokeslets, corresponding physically to point forces), including the method of images and boundary integral methods~\cite{happel2012low,kim2013microhydrodynamics,pozrikidis1992boundary}. The boundary condition on the (rigid) surface of the bacterium is ``no-slip'', i.e.~the flow field matches the velocity of the cell body.
		
	We consider the setup shown in Fig.~\ref{bund:fig:setup}. For simplicity, we assume that the bacterium has a spherical cell body, and is equipped with exactly two rigid helical flagellar filaments, labelled by $i=\{1,2\}$, that point away radially from the centre of the cell body. The filaments are slender rigid left-handed helices, each rotating with the same angular velocity $\bO$ at its base in a counter-clockwise (CCW) fashion (when viewed from behind the cell looking towards it~\cite{berg2008coli}). We non-dimensionalise all lengths so that the cell body has unit radius. The  dimensionless axial length of each flagellar filament is denoted by $L$, and their points of attachment are separated by a solid angle $\theta$ ranging from 0 to $2\pi$. We further assume that the organism is neutrally buoyant. 
		
	Fundamentally, for any rigid body immersed in Stokes flow there is  a linear relationship between its translational and rotational velocities $\bm{U}$ and $\bm{\Omega}$, and the hydrodynamic force $\bF$ and torque $\bG$ that it exerts on the surrounding fluid~\cite{happel2012low}. The so-called resistance matrix instantaneously linking kinematic and dynamic quantities depends only on the fluid viscosity and the geometry (shape and size) of the body. Moreover, for a chiral body it includes terms that couple translation and rotation.
	 
	Consequently, the rotation of each flagellar filaments results in both a force and a torque exerted onto the surrounding fluid~\cite{purcell1997efficiency}. In the case where $\bm{U}_i=\bm{0}$ and $\bO_i$ aligned with the axis of the helix $i$, these can be easily evaluated using resistive-force theory of slender filaments as~\cite{lauga2009hydrodynamics,dauparas2018helical}
	\begin{align}
	\bF_i &= \xi_\perp (1-\tilde{\xi}) \sin\Psi\cos\Psi \rho\Lambda\bm{\Omega}_i,\label{bund:eq:Fi}\\
	\bG_i &= \xi_\perp  (\cos^2\Psi +\tilde{\xi}\sin^2\Psi)\rho^2\Lambda\bm{\Omega}_i,\label{bund:eq:Gi}
	\end{align}
	where $\rho$ is the radius of the flagellar helix, $P$ the helix pitch, $\Psi=\tan^{-1}(2\pi\rho/P)$ the helix pitch angle, $\Lambda=L\sec\Psi$ is the arc-length of the flagellar helix, and $\tilde{\xi}=\xi_{||}/\xi_\perp$ the ratio of { Lighthill's} drag coefficients given by~\cite{lighthill1976flagellar}
	\begin{align}
	\xi_{||}=\frac{2\pi\mu}{\ln\left(1.13\rho \sec\left(\Psi\right)/d\right) },\quad \xi_\perp=\frac{4\pi\mu}{1/2+\ln\left(1.13\rho \sec\left(\Psi\right)/d\right)},
	\end{align}
	which also depend weakly on the filament thickness $d$. { Typically, $d\ll\rho$~\cite{darnton2007torque} and $\tilde{\xi}\approx0.6$}.
	
	 We note that resistive-force theory is only the  leading-order approximation for the hydrodynamic forces in the slenderness of  the filaments. More accurate theories such as slender-body theory~\cite{lighthill96_helical}, or such accounting for elasticity~\cite{katsamba2019propulsion} give more accurate though significantly more complicated estimates for $\bF_i$ and $\bG_i$. However, as we show in \S\ref{bund:sec:thrustcomp}, their absolute values do not actually matter when comparing direct and indirect interactions since they both scale linearly with the kinematic quantities.

	\subsection{Modelling flagellar filaments by flow singularities}
	
	The main idea behind our minimal model is to abstract away the geometric shape of the flagellar filaments by replacing them with their averaged hydrodynamic signatures. Specifically, the effect of each filament on the fluid can be represented by a point force (Stokeslet) and a point torque (rotlet) singularities, located a distance $R=L/2+1$ away from the centre of the cell body and with strengths calculated according to Eqs.~\ref{bund:eq:Fi} and \ref{bund:eq:Gi} respectively. This step is motivated by the well-known multipole expansion for the motion of a rigid body in Stokes flow~\cite{kim2013microhydrodynamics}. For each helical filament, we may write the flow disturbance due to its motion as a series in powers of $1/r$, where $r$ is the distance from a point in the fluid to the centrepoint of the filament. The leading order $\mathcal{O}(1/r)$ term is   equivalent to a point force at the centre of the body, with magnitude equal to the total force it exerts on the surrounding fluid. The next higher order, $\mathcal{O}(1/r^2)$, corresponds to a force dipole, which may be split into symmetric and antisymmetric parts. The antisymmetric part may be interpreted as a point torque, equal to the total torque exerted by the body, while the symmetric part may be calculated as an integral of the stress distribution over its surface~\cite{batchelor1970stress}.
	
	It is known that for a model organism such as \emph{E.~coli} the filaments rotate with a period on the order of $\SI{10}{\milli\second}$, which is much shorter than the time scale of bundle formation, which is around $\SI{500}{\milli\second}$~\cite{turner2000real,darnton2007torque}. We therefore assume that the force $\bm{F}_i$, torque $\bm{G}_i$ and symmetric force dipole that are exerted by flagellum $i$ can be averaged over one period of flagellar rotation and are all oriented approximately in the radial direction. In that case, the force and symmetric force dipole both give rise to an axisymmetric flow field, while the torque decouples and gives an azimuthal flow. We may then discount the symmetric force dipole as it is less dominant at long distances than the  flow induced by the force, but we retain the torque as it is the leading-order contribution in the azimuthal direction. This approach allows us therefore to model the  leading-order flows in both the polar and azimuthal directions. 
	
	Furthermore, the physics of Stokes flow require that the total hydrodynamic force and torque exerted by the swimmer on the fluid vanish~\cite{purcell1977life}, so in the absence of body forces such as gravity, a force $\bF_\text{tot}$ and torque $\bG_\text{tot}$ are also exerted by the cell body on the fluid to maintain dynamic equilibrium. However, as we will show in \S\ref{bund:sec:stokesindirect} and is a subtle point, these do not balance the flagellar thrust and torque exactly, since the flows generated by the actuation of the filaments induce additional stresses and drag on the cell body.
	
	\subsection{Direct vs indirect interactions}\label{bund:sec:definitions}	
	
	With these prerequisites it is now possible to define the main terminology used in our  study. Labelling the positions of the singularities by $\bx_i$, we define the \emph{direct} hydrodynamic interaction  as the magnitude of the flow induced instantaneously at $\bx_2$ (resp.~$\bx_1$) by the singularities located at $\bx_1$ (resp.~$\bx_2$). Since both flagellar filaments are rotated at the same angular velocity $|\bO|$, we have $|\bF_1|=|\bF_2|$ and likewise for the torques, so the magnitude of the direct hydrodynamic interaction is symmetric in the label $i$ used to measure the magnitude of the  flow.	 In contrast, we define the \emph{indirect} interaction as the magnitude of the flow induced instantaneously at $\bx_2$ (resp.~$\bx_1$) by the rigid-body motion of the cell body induced by the hydrodynamic forcing at $\bx_1$ (resp.~$\bx_2$), i.e.~the flow induced by the motion of the cell as measured in the co-moving body frame. 
 	
	In order to illustrate  these definitions intuitively, we can describe two thought experiments highlighting each kind of hydrodynamic interaction separately. For direct interactions consider a bacterium that is held stationary, e.g.~by means of a micro-pipette. In this case the direct hydrodynamic interactions between its flagella are unaffected; both still create forces and torques which move  the  fluid and thus move the neighbouring flagellum.  However, since the cell does not move no flow is created by the motion of the cell body, and  therefore there are no indirect interactions at play. In contrast, to understand indirect interactions imagine that the flagella are shielded from another by a vane-like structure, such as the one featured by the choanoflagellate \emph{Diaphanoeca grandis} \cite{nielsen2017hydrodynamics}. In that case there are no direct interactions since a hydrodynamic forcing contained within one vane does not affect the flow in another; however, as the cell swims the flows induced by the swimming motion advect and rotate the flagella, so indirect interactions are still present.
	
	These two definitions  allow us to characterise a wide range of different hydrodynamic effects,  including that considered in past work. By considering direct interactions, we take into account both the geometric configuration of the flagella relative to one another, as well as any boundary effects due to the no-slip condition on the cell body. This captures both wall-induced attraction~\cite{squires2000like} in the case of the Stokeslet, and wrapping~\cite{man2016hydrodynamic} in the case of the rotlet. Meanwhile, indirect interactions capture global flows such as rotation-induced wrapping~\cite{powers2002role} or zipping~\cite{adhyapak2015zipping} and translation-induced contraction~\cite{riley2018swimming}. 
	
	Furthermore, since all Stokes flows are linear in the force and torque, for any given geometrical setup the ratio of direct to indirect advection is in fact independent of the magnitude of the hydrodynamic forcing, and depends only on  the geometric configuration of the flagellar filaments. In the context of our model, this reduces to the two parameters $R$ (distance to centre of cell body) and $\theta$ (angle between the two flagellar filaments), allowing us to analyse the  parameter space comprehensively. In \S\ref{bund:sec:thrust} we carry out the analysis for the force and consider the torque in \S\ref{bund:sec:rot}.
	
	\section{Advection induced by flagellar thrust}\label{bund:sec:thrust}
	
	\subsection{Direct polar advection between flagella}\label{bund:sec:stokesdirect}
	We first analyse the direct flow due to flagellar thrust. To this end we consider a radial point force $\bF_1$ located at $\bx_1$ and analyse the flow that it generates at location $\bx_2$. For convenience we define Cartesian coordinates so that the cell body is located at the origin and the force $\bF_1$ is aligned with the $z$-axis. The solution to the Stokes equations Eq.~\eqref{bund:eq:Stokes} for a radial point force outside a rigid sphere, originally due to Oseen,  may be written in a compact fashion as~\cite{kim2013microhydrodynamics}
	\begin{align}
	\bm{u}(\bm{x})=\bm{F}_1\cdot\left\{\Oseen(\bm{x}-\bx_1)+\alpha\Oseen(\bm{x}-\bx^*_1)+\beta(\bm{e}\cdot\nabla)\Oseen(\bm{x}-\bx^*_1)+\gamma\nabla^2\Oseen(\bm{x}-\bx^*_1)\right\},\label{bund:eq:radialStokeslet}
	\end{align}
	where we define
	\begin{align}
	\bm{e}&=\bx_1/R,\\
	\bx^*_1&=\bm{e}/R=\bx_1/R^2,\\
	\bm{F}_1 &= F\bm{e},\\
	\alpha &= -\frac{3}{2}R^{-1}+\frac{1}{2}R^{-3},\\
	\beta &=R^{-2}-R^{-4},\\
	\gamma &= -\frac{1}{4}R^{-1}\left(1-R^{-2}\right)^2,\\
	\Oseen(\bm{x}) &= \frac{1}{8\pi\mu}\left(\frac{\bm{I}}{|\bm{x}|}+\frac{\bm{xx}}{|\bm{x}|^3}\right).
	\end{align}
	Here $\Oseen$ is   the Oseen tensor and the point $\bx^*_1$ is the mirror image of $\bx_1$ inside the cell body. In Eq.~\eqref{bund:eq:radialStokeslet}, the quantity $\bF_1\cdot\Oseen$ is the flow due to a point force of strength $\bF_1$ in an unbounded fluid. By taking gradients of $\Oseen$, it is then possible to describe the flow due to force dipoles, quadrupoles and so on. This allows us to interpret the solution in Eq.~\eqref{bund:eq:radialStokeslet} physically in terms of image singularities. The term with coefficient $\alpha$ corresponds to an image force, the $\beta$ term corresponds to a collinear force dipole and the $\gamma$ term is a source dipole (or degenerate force quadrupole). These coefficients are functions of $R$ and ensure that the no-slip boundary condition,
	\begin{align}
	\bm{u}(\bm{x})=\bm{0},\quad\text{if }|\bm{x}|=1,
	\end{align}
	is satisfied at all points on the spherical body. The solution is structurally identical to the image system for a force near a plane wall~\cite{kim2013microhydrodynamics}, albeit with different values for the coefficients. Importantly, the sign of $\alpha$ is negative since $R>1$, and $|\alpha|<1$. Physically this means that an opposite but not equal force needs to be exerted on the cell body to keep it stationary. Balancing this force with drag will give rise to the indirect advection discussed in  \S\ref{bund:sec:stokesindirect}.
	
	\begin{figure}[t]
		\centering
		\begin{subfigure}[b]{.59\linewidth}
			\centering
			\includegraphics[width=\textwidth]{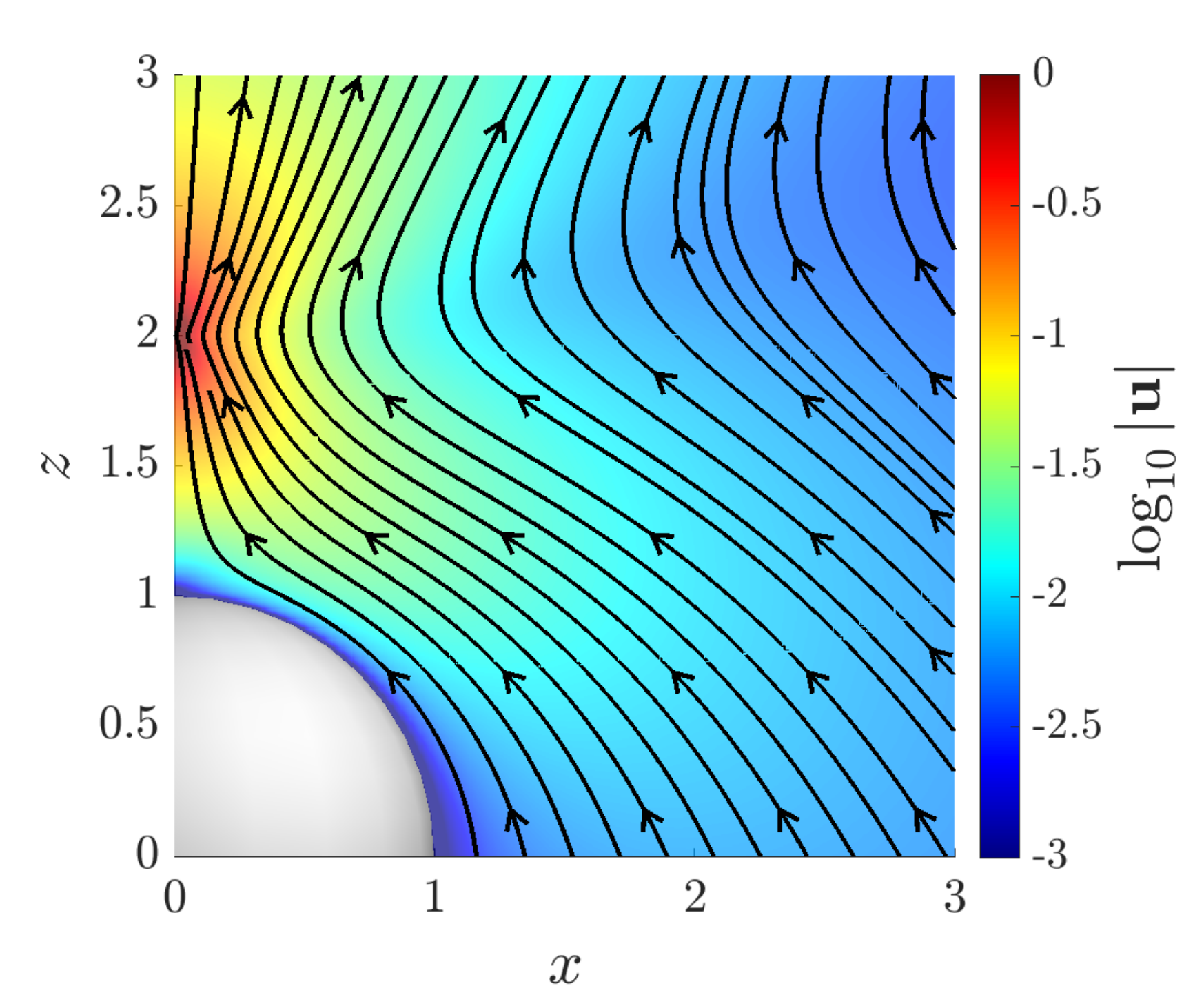}
			\caption{}\label{bund:fig:stokesillust}
		\end{subfigure}
		\hfill
		\begin{subfigure}[b]{.39\linewidth}
			\centering
			\includegraphics[width=\textwidth]{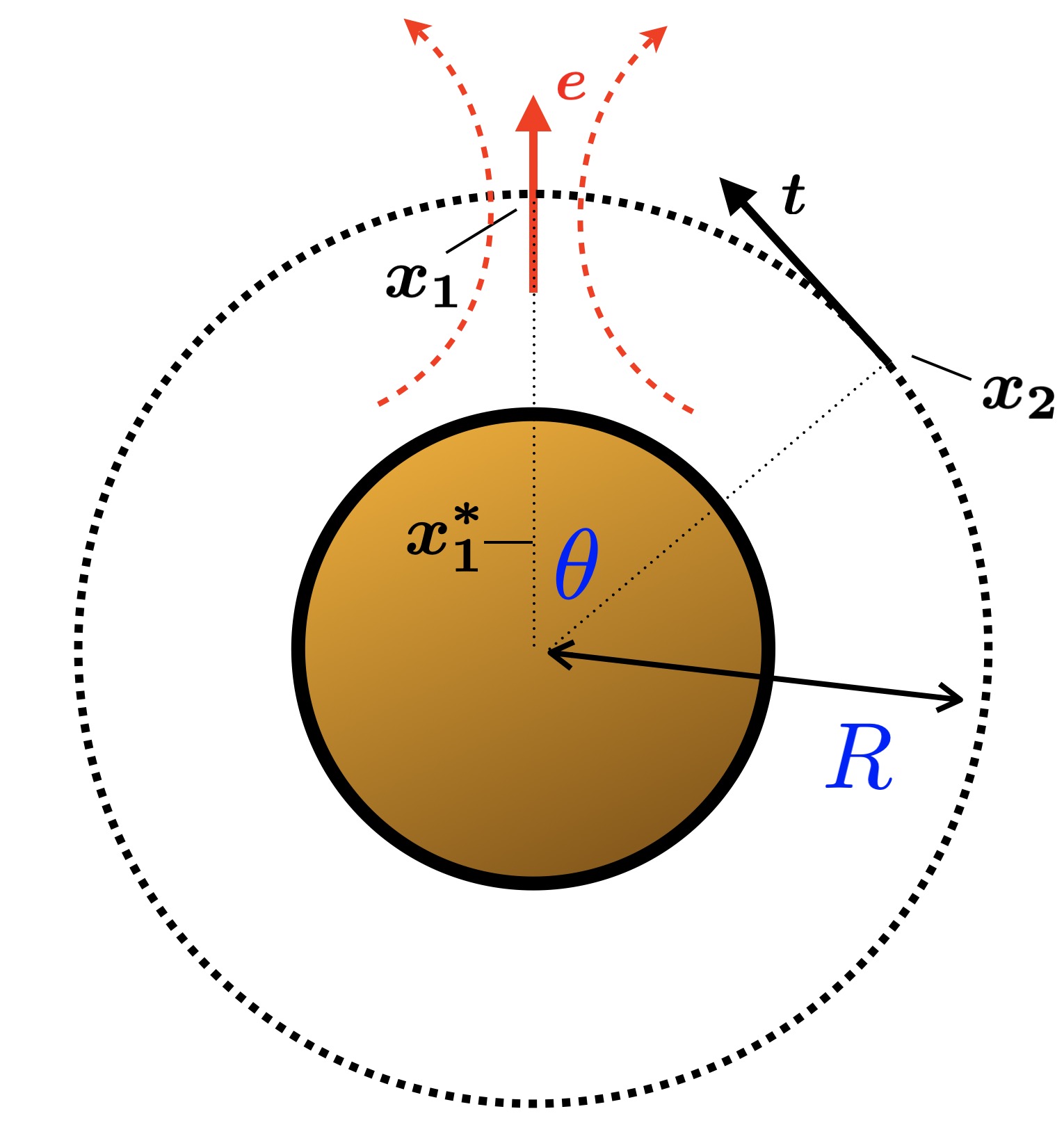}
			\vspace{0pt}
			\caption{}\label{bund:fig:stokessketch}
		\end{subfigure}
		\caption{(a) Illustration of the flow due to a radial point force with $R=2$ and $\bF=\hat{\bm{z}}$ in a Cartesian coordinate system with the cell body centred at the origin in the bottom-left corner. Colours quantify iso-magnitudes of the  flow  and black streamlines are superimposed.  The force induces a flow that advects fluid around the cell body towards the singularity. (b) Sketch of the relevant geometric quantities to calculate the direct polar advection induced at $\bx_2$ by a point force at $\bx_1$ (see text for notation).}
	\end{figure}
	
	We illustrate the flow induced by such a radial force in the case $R=2$ in Fig.~\ref{bund:fig:stokesillust}. The effect of the cell body is a distortion of the streamlines leading to advection predominantly in the polar $\theta$-direction. Since we are interested in the advection of an inextensible flagellum of the same length, we focus therefore on the polar component of the flow. We write $\bx_2=R(\sin\theta,0,\cos\theta)$ and define $\bm{r}=\bx_2-\bx_1$,  and $r=|\bm{r}|$ as the distance to the singularity. In a similar fashion we use $\bm{r}^*=\bx_2-\bx_1^*$ and $r^*=|\bx_2-\bx_1^*|$ to describe the distance to the image point. Both $r$ and $r^*$ may be written in terms of $R$ and $\theta$ as
	\begin{equation}
	r= 2R\sin\frac{\theta}{2},\quad r^*= \sqrt{\left(R-R^{-1}\right)^2+4\sin^2\frac{\theta}{2}}.
	\end{equation}
	Finally we define the vector $\bm{t}=(-\cos\theta,0,\sin\theta)$. An illustrative sketch is provided in Fig.~\ref{bund:fig:stokessketch}.
	
	The flow component of interest for direct interactions, which we call $u^\text{dir}_{F}$, is then given by
	\begin{align}\label{bund:eq:ulat_A}
	u^\text{dir}_{F}&\equiv\bm{t}\cdot\bm{u}(\bx_2)\nonumber\\
	&=\frac{F}{8\pi\mu}\cdot\left\{\bm{e}\cdot\Oseen(\bm{r})\cdot\bm{t}+\alpha\bm{e}\cdot\Oseen(\bm{r}^*)\cdot\bm{t}+\beta\bm{e}\cdot(\bm{e}\cdot\nabla)\Oseen(\bm{r}^*)\cdot\bm{t}+\gamma\bm{e}\cdot\nabla^2\Oseen(\bm{r}^*)\cdot\bm{t}\right\}\nonumber\\
	&=\frac{3F}{16\pi\mu}\left[\frac{1}{r}-\frac{1}{R}\frac{1}{r^*}-\frac{\left(R^2-1\right)^2}{2R^3}\frac{1}{r^{*3}}+\frac{\left(R^2+1\right)\left(R^2-1\right)^3}{2R^5}\frac{1}{r^{*5}}\right]\sin\theta.
	\end{align}
	The term in brackets is a complicated function of $R$ and $\theta$ but strictly positive for $R>1$ and $0 < \theta < \pi$, i.e.~for every possible geometric configuration. The generation of thrust by each flagellar filament is therefore sufficient to facilitate the polar attraction of neighbouring filaments from direct hydrodynamic interactions,  eventually leading to the formation of a bundle.
	
	\subsection{Cell kinematics and indirect polar advection between flagella}\label{bund:sec:stokesindirect}
	
	We now consider the flow induced in the cell body frame due to its swimming motion. In order to find the translation velocity $\bU$ we exploit the fact that the swimmer as a whole is force-free~\cite{purcell1977life}. Neglecting the viscous drag on the flagellar filaments, we have the classical Stokes flow result $\bU=\bF_\text{tot}/6\pi\mu$~\cite{kim2013microhydrodynamics} where the total force $\bF_\text{tot}$ that the cell body exerts on the fluid is given by
	\begin{align}\label{bund:eq:Ftot}
	\bm{F}_\text{tot}=-\sum_i (1+\alpha) \bm{F}_i.
	\end{align}
	The factor $(1+\alpha)$ arises because the thrust that the flagellar filaments exert is already partially balanced by the hydrodynamic image system described in \S\ref{bund:sec:stokesdirect}. Physically, the fact that  $\alpha\neq0$ may be attributed to the flagellar filament-induced flow that entrains the cell body   towards the  flagellum, thus increasing drag in the direction of motion.
	
	For $\bx_1$ and $\bx_2$ as defined in \S\ref{bund:sec:stokesdirect} the total force in Eq.~\eqref{bund:eq:Ftot} evaluates to
	\begin{align}
	\bF_\text{tot}=2F(1+\alpha)\cos\frac{\theta}{2}\left(-\sin\frac{\theta}{2},0,-\cos\frac{\theta}{2}\right),
	\end{align}
	from which it follows that the swimming  speed is
	\begin{align}
	|\bU|=\frac{2F}{6\pi\mu}\left(1-\frac{3}{2}R^{-1}+\frac{1}{2}R^{-3}\right)\cos\frac{\theta}{2}.
	\end{align}
	{ It is important to note here that, in general, this does not give an explicit expression for $\bU$ since $F$ itself may depend on $\bU$ and not just $\bO_i$ as postulated by Eq.~\eqref{bund:eq:Fi}. However, as shown in \S\ref{bund:sec:thrustcomp} below, the ratio of direct to indirect flows is independent of $F$, so this does not affect the conclusions we can draw from our model.}
	
	The flow field in the body frame is then   given by the classical expression for the flow around a translating rigid sphere, which is~\cite{kim2003macroscopic}
	\begin{align}
	\bm{u}(\bm{x})&=\frac{\bm{F}_\text{tot}}{8\pi\mu}\cdot \left\{ \Oseen(\bm{x}) +\frac{1}{6}\nabla^2\Oseen(\bm{x})\right\} -\frac{\bm{F}_\text{tot}}{6\pi\mu}\nonumber\\
	&=\frac{\bm{F}_\text{tot}}{8\pi\mu}\left(R^{-1}+\frac{1}{3}R^{-3}-\frac{4}{3}\right)+\frac{\bm{F}_\text{tot}\cdot\bm{x}\bm{x}}{8\pi\mu}\left(R^{-3}-R^{-5}\right).
	\end{align}
	From this it is straightforward to find the indirect advective polar component, $u^\text{ind}_{F}$, which is given by
	\begin{align}\label{bund:eq:ulat_P}
	u^\text{ind}_{F}\equiv\bm{t}\cdot \bm{u}(\bm{x}_2)= \frac{F}{6\pi\mu}\left(1-\frac{9}{4}R^{-1}+\frac{9}{8}R^{-2}+\frac{1}{4}R^{-3}-\frac{1}{8}R^{-6}\right)\sin\theta.
	\end{align}
	Notably,   unlike the direct advection from Eq.~\eqref{bund:eq:ulat_A}, the indirect flow is simply proportional to $\sin\theta$. It is also strictly positive in the entire parameter range, and thus it also always contributes to bundling.
	
	\subsection{Direct vs indirect interactions}\label{bund:sec:thrustcomp}
	From Eqs.~\eqref{bund:eq:ulat_A} and \eqref{bund:eq:ulat_P} we see that both direct and indirect advection always facilitate bundling between two flagellar filaments, since the induced lateral velocity is positive in both cases. The fundamental question is therefore under which condition one type  of hydrodynamic interaction dominates the other. To this end we consider their ratio, which is  given by
	\begin{equation}\label{bund:eq:Frat} 
		\frac{u^\text{dir}_{F}}{u^\text{ind}_{F}}=\frac{9}{8}\frac{r^{-1}-\left(Rr^*\right)^{-1}-\frac{1}{2}\left(R^2-1\right)^2\left(Rr^*\right)^{-3}+\frac{1}{2}\left(R^2+1\right)\left(R^2-1\right)^3\left(Rr^*\right)^{-5}}{1-\frac{9}{4}R^{-1}+\frac{9}{8}R^{-2}+\frac{1}{4}R^{-3}-\frac{1}{8}R^{-6}}.
	\end{equation}
	Since both $u^\text{dir}_{F}$ and $u^\text{ind}_{F}$ are linear in $F$, this ratio is independent of the flagellar forcing and hence only a function of the geometric parameters $R$ and $\theta$. Physically, this means that the relative contribution to bundling is independent of the (identical) rotation rate of the filaments. Only the absolute magnitude of the advection and hence the time scale on which bundling takes place depend (linearly) on the strength of the actuation, as expected for Stokes flow~\cite{happel2012low}.
		
	\begin{figure}[t]
		\centering
		\includegraphics[width=0.7\textwidth]{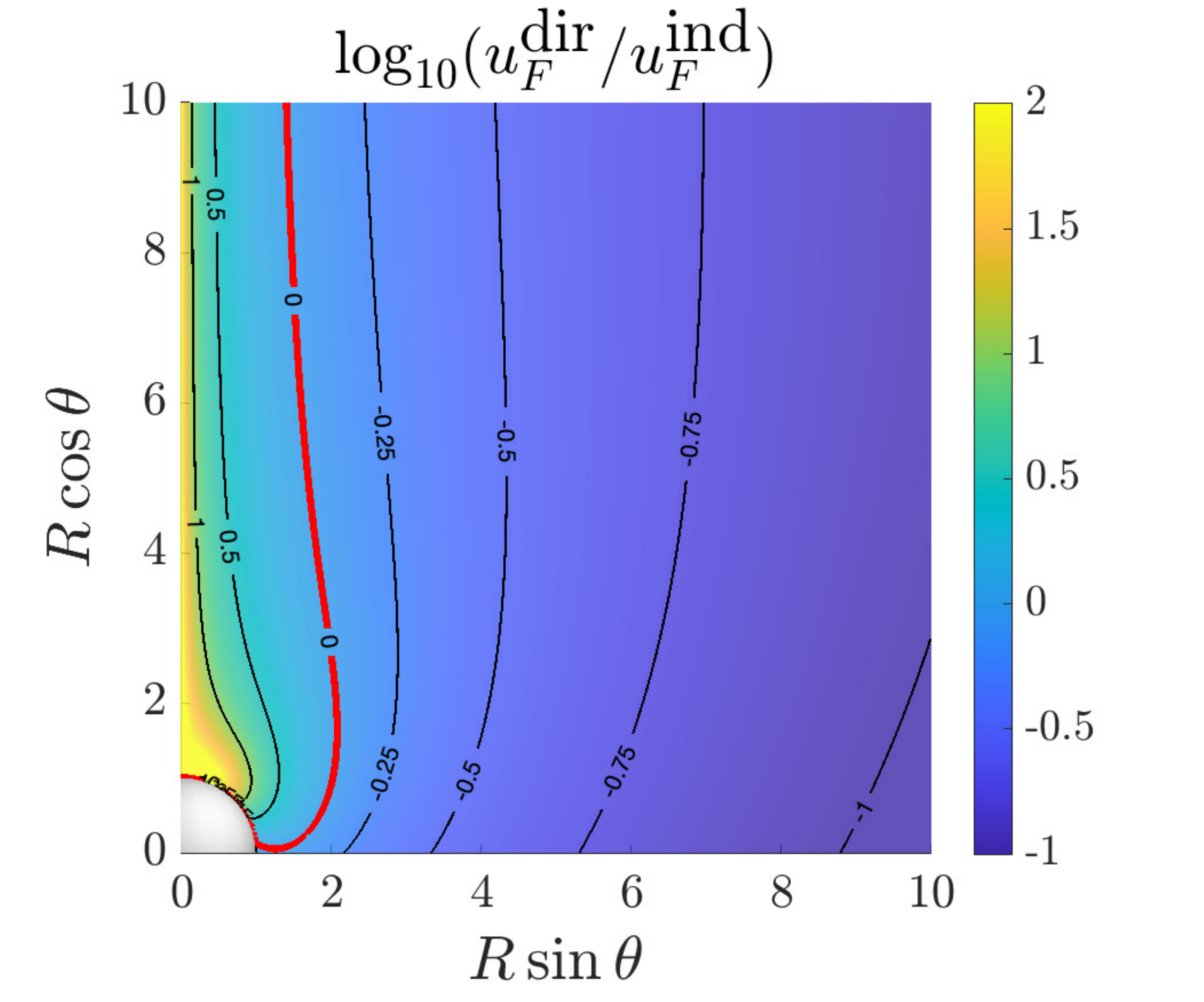}
		\caption{Contour plot of the ratio of direct to indirect polar advection, $	u^\text{dir}_{F}/	u^\text{ind}_{F}$, as induced by flagellar thrust. The magnitude is displayed using a logarithmic scale. The thick red line indicates the boundary where direct and indirect  interactions exactly balance ($	u^\text{dir}_{F}/	u^\text{ind}_{F}=1$). At close separation and for short filaments direct interactions dominate, while for large separation and long filaments the balance is reversed.}\label{bund:fig:Frat}
	\end{figure}
	
	  In order to compare $u^\text{dir}_{F}$ to $u^\text{ind}_{F}$, we display the ratio $u^\text{dir}_{F}/u^\text{ind}_{F}$ graphically in Fig.~\ref{bund:fig:Frat} using a logarithmic scale. The diagram is to be interpreted as follows: one flagellar filament is located on the vertical axis, with a flow singularity at its midpoint, at a distance $R$ from the origin; the colour of a point in the plane then indicates the ratio of direct to indirect advection that would be experienced by the midpoint of the second flagellar filament  (i.e.~a flow singularity) located at that position. Points on the vertical axis therefore correspond to two flow singularities on top of each other, while points on the horizontal axis correspond to filaments at right angles, $\theta=\pi/2$. 
	
	The thick red solid line in in Fig.~\ref{bund:fig:Frat} indicates the contour on which direct and indirect effects balance exactly, i.e.~$u^\text{dir}_{F}/u^\text{ind}_{F}=1$. From Eq.~\eqref{bund:eq:Frat} we deduce that this line behaves asymptotically as
	\begin{align}
		R\sin\theta\to\frac{9}{8},\quad\text{as }R\to\infty,
	\end{align}
	creating a closed wake-like region surrounding the vertical axis inside of which direct interactions dominate. This observation may be rationalised  physically. The indirect interactions are proportional to $\sin\theta$, so when the two filaments  are nearly aligned then   the flow generated by translation of the cell body only has a small component in the polar direction that brings them closer together.  In contrast, direct hydrodynamic interactions grow when the filament-filament separation decreases; this is true in the singular representation used here and is also true for real finite-sized filaments. Direct effects dominate therefore at close separation between the filaments, while passive effects dominate at wide separations. Thinking in terms of a dynamic bundling process, we therefore obtain  that indirect effects dominate at early stages, and are then surpassed by the direct interactions when the separation between the filaments decreases and they enter each other's `wake'.
	
	Another important conclusion obtained  from Fig.~\ref{bund:fig:Frat} is that indirect effects are more important for large $R$, i.e.~for long flagellar filaments relative to the cell body. Indeed, for a fixed value of $R$, we see that a larger proportion of the $\theta$-range is indirect-dominated. Furthermore, as $R\to\infty$ with the separation angle $\theta$ fixed, the direct advection decays to zero, while the indirect component tends to a finite value, $\tfrac{1}{2}\sin\tfrac{\theta}{2}|\bm{U}|$, because the filaments are moving in the body frame of the cell, whose swimming speed asymptotes to a  finite value. The converse is true in the limit of short filaments, i.e.~$R\to1$, where direct interactions always dominate. This is because the swimming speed disappears, $|\bm{U}|\to0$, as $R\to1$, due to the backwards drag that short flagella exert on the cell body (i.e.~$\alpha\to-1$ in Eq.~\ref{bund:eq:Ftot}).

 	\section{Advection induced by flagellar rotation}\label{bund:sec:rot}
 	
	In section \S\ref{bund:sec:thrust} we analysed the effect of flagellar thrust on bundling by modelling individual filament as radial Stokeslets. Here we conduct a similar analysis for the flow induced by flagellar rotation and therefore focus on the flows created by rotlets, i.e.~point torques. While thrust generates both direct and indirect polar flows that act to reduce the separation between flagellar filaments, we show here that the flows created by rotation only have components in the azimuthal direction, and therefore lead only to relative rotation but not, on their own, to a relative change in the separation between the filaments.
	
	The expression for a radial rotlet located outside a rigid sphere was recently derived in Ref.~\cite{chamolly2020stokes}. The flow field can be written as
	\begin{align}
		\bm{u}(\bx)&=\bG_1\times\left[\bm{\mathcal{R}}(\bx-\bx_1)+\delta \bm{\mathcal{R}}(\bx-\bx_1^*)\right],\\
		\bG_1&=G\bm{e},\\
		\delta&=-R^{-3},\\
		\bm{\mathcal{R}}(\bx)&=-\frac{1}{8\pi\mu}\bn\frac{1}{|\bx|},
	\end{align}
	and the image system may be interpreted as simply due to a single rotlet located at the mirror image point $\bx_1^*$. As claimed, the only non-zero component of $\bu$ is in the azimuthal $\phi$-direction, decoupling it therefore from advection due to thrust. This direct component, which we denote by $u_G^\text{dir}$, is given by
	\begin{align}\label{eq:udirG}
		u_G^\text{dir}\equiv\hat{\bm{\phi}}\cdot\bm{u}(\bx_2)=\frac{GR}{8\pi\mu}\left[\frac{1}{r^3}-\frac{1}{\left(Rr^*\right)^3}\right]\sin\theta,
	\end{align}
	where $\hat{\bm{\phi}}=(0,1,0)$. 
	
	In order to compute the indirect interactions, we need to calculate the torque $\bm{G}_\text{tot}$ exerted by the cell body on the fluid. In analogy with Eq.~\eqref{bund:eq:Ftot}  it is
	\begin{align}\label{bund:eq:Gtot}
		\bm{G}_\text{tot}&=-\sum_i (1+\delta) \bm{G}_i\nonumber\\
		&=2G(1+\delta)\cos\frac{\theta}{2}\left(-\sin\frac{\theta}{2},0,-\cos\frac{\theta}{2}\right),
	\end{align}
	and $-1<\delta<0$ for $R>1$. Neglecting the viscous drag on the filaments, the rotation rate of the cell body is then found to have magnitude
	\begin{align}
		|\bm{\Omega}|=\frac{G}{4\pi\mu}\left(1-R^{-3}\right)\cos\frac{\theta}{2},
	\end{align}
	where we used the well-known value for the rotational drag on a unit sphere, $\bm{G}=-8\pi\mu\bm{\Omega}$~\cite{kim2013microhydrodynamics}. The indirect swirling flow experienced in the body frame is then found from the classical formula for the flow due to a rotating rigid sphere,
	\begin{align}
	\bm{u}(\bm{x})&=\frac{\bm{G}_\text{tot}}{8\pi\mu}\times \left\{\bm{\mathcal{R}}(\bm{x}) -\bm{x} \right\},
	\end{align}
	which again only has an azimuthal component,
	\begin{align}\label{eq:uindG}
	u^\text{ind}_G\equiv\hat{\bm{\phi}}\cdot\bu(\bx)=\frac{GR}{8\pi\mu}\left(1-R^{-3}\right)^2\sin\theta.
	\end{align}

	\begin{figure}[t]
		\centering
		\includegraphics[width=0.7\textwidth]{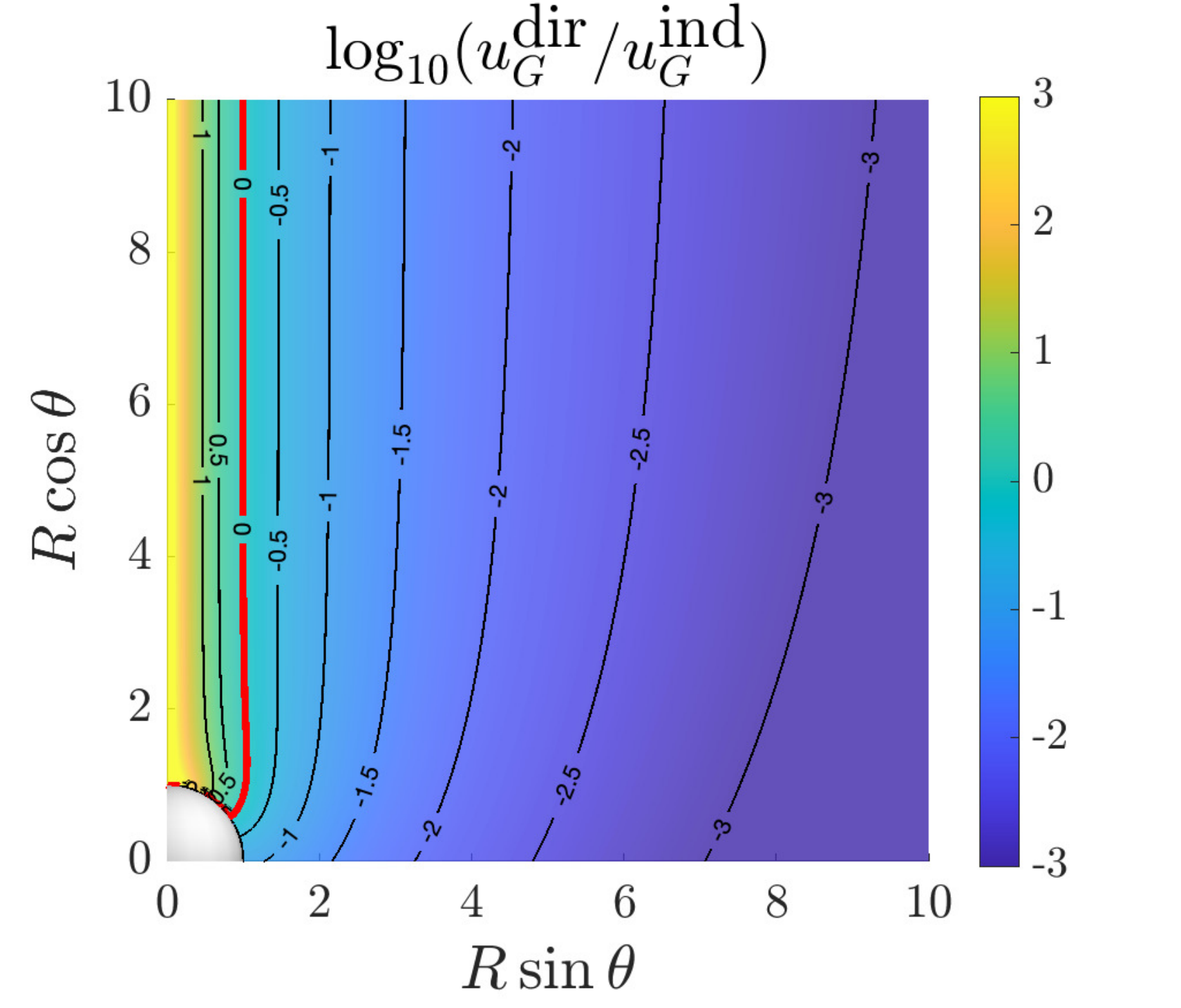}
		\caption{Contour plot of the ratio between direct and indirect azimuthal advection, ${u^\text{dir}_G}/{u^\text{ind}_G}$,  due to flagellar rotation (logarithmic scale). The thick red contour indicates the boundary where direct and indirect interactions exactly balance. As for the thrust, direct rotation dominate at close separation of the filaments, while indirect rotation dominates for large separation.}\label{bund:fig:Grat}
	\end{figure}
		
	Hence the ratio between direct (Eq.~\ref{eq:udirG}) and indirect (Eq.~\ref{eq:uindG}) rotational advection is given by 
	\begin{align}\label{bund:eq:Grat}
		\frac{u^\text{dir}_G}{u^\text{ind}_G}=\frac{r^{-3}-\left(Rr^*\right)^{-3}}{\left(1-R^{-3}\right)^2}.
	\end{align}
	In Fig.~\ref{bund:fig:Grat} we illustrate this ratio as we did for the thrust in \S\ref{bund:sec:thrustcomp}  (Fig.~\ref{bund:fig:Frat}). The result is very similar in its general shape and features to, but there are a couple of notable differences. The region of direct dominance is smaller, with Eq.~\eqref{bund:eq:Grat} implying that
	\begin{align}
	R\sin\theta\to 1,\quad\text{as }R\to\infty,
	\end{align}
	and we also note a smaller bulge of the dividing contour for small values of the flagellar length $R$. This is for two reasons. The first is that the direct rotational interactions decay faster, since the rotlet decays as $1/r^2$ while the Stokeslet (which is absent here) decays as $1/r$. The second is that hydrodynamic interactions with the cell body are less significant, and therefore for short filaments, i.e.~small  values of $R$, the indirect interactions are stronger. Another difference is that, in the case of rotation, the dominance of indirect advection is generally stronger in magnitude (compare scales for the colour scheme in Figs.~\ref{bund:fig:Frat} and \ref{bund:fig:Grat}). This is again due to the faster spatial decay of the rotlet flow.
		
	\section{Dynamic elastohydrodynamic model}\label{bund:sec:numerics}
	
	\subsection{Motivation}
	While  the minimal model allows us to calculate and compare hydrodynamic interactions exactly, it can only do so under very simplified modelling assumptions. In particular, it does not take into account the fact that  flagellar filaments are flexible and tethered at  their base to a fixed location on the cell body. It is also limited by the assumption that the filaments point radially outward, and only applies instantaneously. In order to address these shortcomings, we complement our theory with a numerical elastohydrodynamic model and use it to simulate the bundling dynamics of the helical axes of two inextensible flexible flagellar filaments.

	Numerous sophisticated computational models have been proposed to tackle the bundling of bacterial flagella \cite{adhyapak2015zipping,kanehl2014fluid,flores2005study}. These numerical approaches usually employ variants of the boundary element method and slender-body theory, resolving the structure of the flagellar helix exactly and obtaining accurate approximations to cellular dynamics. However, by calculating flow velocities using such realistic computational method, some of the physical origin of the induced flows can be difficult to grasp. Here, instead of  attempting to model the details of the flagellar filament geometry, we follow an approach similar to past work on the rheology and dynamics of fibres and polymers, which models each flagellum as a rod-spring chain~\cite{reigh2012synchronization,cruz2012review,watari2010hydrodynamics,gebremichael2006mesoscopic,jendrejack2002stochastic,janssen2011coexistence,schmid2000simulations}. Specifically, we propose a rod-spring singularity model that describes the dynamics of the {\it helical axis} of each  flagellar filament. After summarising the model setup in \S\ref{bund:sec:nummodel}, we use it in \S\ref{bund:sec:numresults} to quantify the individual contributions of direct and indirect hydrodynamic interactions during the dynamic bundling process. 
	
	\subsection{Computational modelling}\label{bund:sec:nummodel}
	
	\begin{figure}[t]
		\centering
		\includegraphics[width=0.8\textwidth]{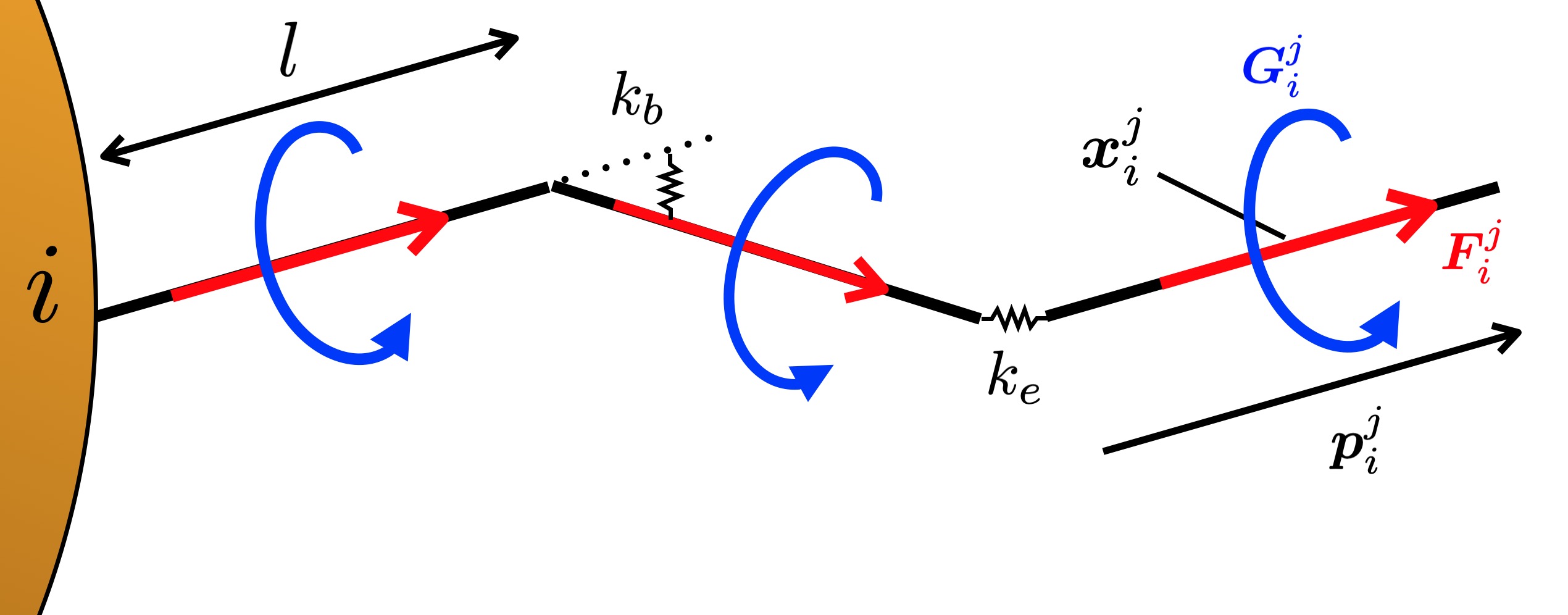}
		\caption{Sketch of the computational model. The helical axis of each flagellar filament, labelled by $i$, is divided into straight rod segments labelled by $j$, each of length $l$, orientation $\bp_i^j$ (unit vectors), and centred at $\bx_i^j$. A force $\bF_i^j$ and torque $\bG_i^j$, each parallel to $\bp_i^j$, act on  the fluid at location $\bx_i^j$. Segments are connected by extensional springs of stiffness $k_e$ and torsional springs of stiffness $k_b$.}\label{bund:fig:numsketch}
	\end{figure}
	
	An illustration of our computational model is shown in Fig.~\ref{bund:fig:numsketch}. We divide the helical axis of the $i$th flagellar filament into $N$ straight segments that we label by $j=1,\dots,N$. We call the outward pointing unit tangent vector to each segment $\bm{p}_i^j$. At the centre point of each segment $\bx_i^j$ we place a point force $\bF_{i}^j$ and a point torque $\bG_i^j$ that are both parallel to, and pointing in the same direction as, $\bm{p}_i^j$ and have magnitudes $|\bF_{i}^j|=|\bF_{i}|/N$ and $|\bG_{i}^j|=|\bG_{i}|/N$ respectively. The first segment is tethered to a point on the cell body, and the segments are linked by extensional and torsional springs of stiffness $k_e$ and $k_b$, respectively. Since the axial length of each filament is $L$, each segment has length $l=L/N$. The cell body has unit radius as in the minimal model. In contrast however, we now allow the forces and torques to have an arbitrary orientation, since $\bp_i^j$ need not point in the radial direction at all times. For our numerical calculations we use the analytical expressions for the corresponding image systems and flows that are available in Ref.~\cite{kim2013microhydrodynamics} for the Stokeslet and in Ref.~\cite{chamolly2020stokes} for the rotlet.
	
	For each filament (omitting the index $i$ for clarity) the set of $2N$ vectors $\{\bx^j,\bm{p}^j\}$ is subject to the evolution equations
	\begin{align}
		\dot{\bx}^j&=\sum_{n=1}^{N}\bu^{\text{dir},n}_{k}\left(\bx^j\right) + \bu^\text{ind}\left(\bx^j\right)+\mathbb{A}^{-1,j}\cdot\left(\bf_\text{ex}^{j+1}-\bf_\text{ex}^{j}+\bf_\text{ster}^{j}\right),\label{bund:eq:xevo}\\
		\dot{\bp}^j&=\bm{\omega}^j\times\bp^j,
	\end{align}
	where
	\begin{align}
		\bm{\omega}^j=\mathbb{C}^{-1,j}\cdot\left[\frac{l}{2}\bp^j\times\left(\bf_\text{ex}^{j+1}+\bf_\text{ex}^{j}\right)+h^j\bp^j\times\bf_\text{ster}^{j}+\bm{g}^{j+1}_{\text{bend}}-\bm{g}^{j}_{\text{bend}} \right].\label{bund:eq:pevo}
	\end{align}
	The first term on the right-hand side of Eq.~\eqref{bund:eq:xevo} describes advection due to direct hydrodynamic interactions. Here $k\neq i$ labels the second filament, so that direct interactions with the individual forces and torques from the other filament are included, but self-interactions are neglected. 
	
	The second term  on the right-hand side of Eq.~\eqref{bund:eq:xevo} describes advection due to indirect interactions. These are found by first solving for the swimming kinematics of the cell body (as we did for the minimal model) and then applying the classical solution for flow due to a translating and rotating rigid sphere~\cite{kim2013microhydrodynamics}. For this, we again need to find the total hydrodynamic force $\bF_\text{tot}$ and torque $\bG_\text{tot}$ exerted by the cell body onto the fluid. Using expressions from Refs.~\cite{kim2013microhydrodynamics,chamolly2020stokes} for the hydrodynamic interactions with the cell body these become
	\begin{align}
		\bF_\text{tot}&=\sum_{i,j} \left[\left(\frac{3}{4R_i^j}+\frac{1}{4{R_i^j}^{3}}-1\right)\bF_i^j
		+\left(\frac{3}{4R_i^j}-\frac{3}{4{R_i^j}^{3}}\right)\bF_i^j\cdot\hat{\bm{r}}_i^j\hat{\bm{r}}_i^j
		-\frac{3}{4{R_i^j}^{2}}\bG_i^j\times\hat{\bm{r}}_i^j\right],\\
		\bG_\text{tot}&=\sum_{i,j}\left[\left(-\frac{1}{2{R_i^j}^{3}}-1\right)\bG_i^j
		+\frac{3}{2{R_i^j}^{3}}\bG_i^j\cdot\hat{\bm{r}}_i^j\hat{\bm{r}}_i^j
		+\left({R_i^j}-\frac{1}{{R_i^j}^{2}}\right)\bF_i^j\times\hat{\bm{r}}_i^j\right],
	\end{align}
	where $\bx_i^j=R_i^j\hat{\bm{r}}_i^j$.
	
	
	The remaining terms in Eqs.~\eqref{bund:eq:xevo} and \eqref{bund:eq:pevo} describe dynamic forces due to extension, steric interactions and bending. The tensors $\mathbb{A}^{-1,j}$ and $\mathbb{C}^{-1,j}$ are hydrodynamic motility tensors for the segments, which are approximated  by the corresponding expressions for prolate ellipsoids~\cite{kim2013microhydrodynamics} with a short semi-axis length equal to $r$ (the filament radius) and the long semi-axis length equal to $l/2$ (the segment length). The long axis is parallel to $\bp^j$ by definition, which introduces the explicit dependence on $j$.
	
	The extensional restoring force exerted on the $(j-1)$th segment by the hinge linking it to the $j$th segment, $\bf_\text{ex}^{j}$, is given by
	\begin{align}
		\bf_\text{ex}^{j}=k_e\left[\bx^j-\bx^{j-1}-\frac{l}{2}\left(\bp^j+\bp^{j-1}\right)\right],
	\end{align}
	where $k_e$ is the strength of a Hookean elastic spring. Since real flagella are almost inextensible, we   choose $k_e$ sufficiently large to suppress any significant extension of the filament.
	
 	Steric interactions are often included in elastohydrodynamic models to avoid the overlapping of physical filaments. In our case they serve the additional purpose of maintaining a minimum distance between individual singularities. Since the singularities are located on the helical axis of the flagellar filaments, we   use these steric interactions to prevent an approach of less than $\rho$, the radius of the flagellar helix (i.e.~the radius of the  cylinder on which the helix is coiled). Specifically, we choose
	\begin{align}
		\bf_\text{ster,i}^{j}=k_s\exp\left(-20\frac{|\bd_{ik}^j|-\rho}{\rho}\right)\frac{\bd_{ik}^j}{|\bd_{ik}^j|},
	\end{align}
	where $k_s$ is the strength of the interaction, and
	\begin{align}
		\bd_{ik}^j=\bx_i^j+h_i^j\bp_i^j-\bx_k^j+h_k^j\bp_k^j,
	\end{align}
	and
	\begin{align}
		h_i^j=\frac{\left(\bx_k^j-\bx_i^j\right)\cdot\left(\bm{I}-\bp_k^j\bp_k^j\right)\cdot\bp_i^j}{1-\left(\bp_i^j\cdot\bp_k^j\right)^2},\quad 		h_k^j=\frac{\left(\bx_i^j-\bx_k^j\right)\cdot\left(\bm{I}-\bp_i^j\bp_i^j\right)\cdot\bp_k^j}{1-\left(\bp_i^j\cdot\bp_k^j\right)^2},
	\end{align}
	are the distances from the midpoint of the $\{i,j\}$-th and $\{k,j\}$-th segment, respectively, where the steric forces act. Since the segments have a finite length $l$, the force is set to zero if $|h_i^j|>l/2$ or $|h_k^j|>l/2$. By symmetry of the dynamics, only $N$ different interactions are possible and need to be calculated.
	
	Finally, the bending torque exerted on the $(j-1)$th segment by the hinge connecting to the $j$th segment, $\bm{g}_\text{bend}^{j}$, is included as
	\begin{align}
	\bm{g}_\text{bend}^{j}=k_b\bp^{j-1}\times\bp^j,
	\end{align}
	where $k_b$ is a torsional spring, related to the bending modulus of the filament, $EI$, as	$k_b=EI/l$. At the base where the more flexible hook is located, we choose $k_{b,\text{hook}}=k_b/20$ in agreement with Ref.~\cite{janssen2011coexistence}.
	
	Two physical contributions are absent from Eq.~\eqref{bund:eq:pevo}, namely vorticity and torsion. Vorticity influences the rotation of filament segments but neglecting it is justified by the stiffness and inextensibility of the filament, since the strengths of the extensional and torsional springs $k_e$ and $k_b$ are sufficiently large to dominate the dynamics. The torsion term is neglected following the scaling argument presented in Ref.~\cite{powers2002role}, which shows that torsion is insignificant compared to extension and bending.
	
	\begin{table}[t]
		\centering
		\begin{tabular}{c|c|c|c}
			Parameter & Symbol & Value & Source \\
			\hline
			\hline
			Fluid viscosity & $\mu$ & $\SI{1.00}{\milli\Pa\second}$ &\cite{haynes2014crc} \\
			Cell body radius &   & $\SI{1}{\micro\metre}$ & \cite{darnton2007torque}, see text. \\
			Bending stiffness & $EI$ & $\SI{3.5}{\pico\N\micro\metre\squared}$  &\cite{darnton2007torque} \\
			Helix radius & $\rho$ & $\SI{0.2}{\micro\metre}$ &\cite{darnton2007torque} \\
			Helix pitch & $P$ & $\SI{2.22}{\micro\metre}$ &\cite{darnton2007torque} \\
			Filament thickness & $d$ & $\SI{12}{\nano\metre}$ &\cite{darnton2007torque} \\
			Filament rotation rate & $|\bm{\Omega}_i|$ & $\SI{697}{\per\second}$ &\cite{darnton2007torque} \\
			Thrust per length & $|\bF_{i}|/L$ & \quad $\SI{89.9}{\femto\N\per\micro\metre}$ \quad& Eq.~\eqref{bund:eq:Fi}\\
			Torque per length & $\quad|\bG_{i}|/L \quad$ & $\SI{87.0}{\femto\N}$ & Eq.~\eqref{bund:eq:Gi} \\
			\hline
			\# of segments per length & $N/L$  & $\SI{2}{\per\micro\metre}$ & N/A \\
			Steric force constant & $k_s$ & $\SI{400}{\pico\N}$ & N/A \\
			Extensional spring constant \quad &  $k_e$  & $\SI{100}{\pico\N\per\micro\metre}$ & N/A \\
			Flagellar axial length & $L$ & ${4,6,8}~\si{\micro\metre}$ & \quad \cite{darnton2007torque}, see text. \\
			Base separation angle & $\theta$ & \quad${\pi}/{8},{\pi}/{4},
			{3\pi}/{8}, {\pi}/{2}$\quad  & N/A \\
		\end{tabular}
		\vspace{12pt}
		\caption{Experimentally-obtained physical parameters for our computational elastohydrodynamic model as applicable to the swimming  of \emph{E.~coli}. We assume that the motor rotation rate $|\bm{\Omega}_i|$, rather than the motor torque $|\bG_{i}|$, is held constant in order to compare filaments of different lengths $l$.}\label{tablelabel}
	\end{table}
	
	
	\subsection{Numerically simulated filament dynamics}\label{bund:sec:numresults}
	
	Our computational model has  a large number of  parameters, so we take representative values for many of them from the experimental literature. We focus on the model bacterium \emph{E.~coli}, and summarise these values  in Table~\ref{tablelabel}. We note that there exists considerable spread in the bending moduli of flagellar filaments across species~\cite{kim2005deformation,katsamba2019propulsion}, as well as specifically for \emph{E.~coli} in the flagellar length~\cite{turner2012growth} and estimates for the motor torque~\cite{das2018computing}. Nevertheless, these representative values are sufficient to gain an understanding of the general dynamics of bundling.
	{ Furthermore, the strengths of the singularities are calculated using Eqs.~\eqref{bund:eq:Fi} and \eqref{bund:eq:Gi}, i.e.~using resistive-force theory without taking into account the \emph{a priori} unknown translational velocity of the flagellar filaments, or hydrodynamic interactions with the cell body. While our computational model does take into account the viscous drag on the flagellar filaments for the purpose of calculating their dynamics, we acknowledge that the error in the forcing may have a small influence our results.}
	
	We also make the simplifying assumption that the cell body is spherical. In reality, \emph{E.~coli} resembles a prolate ellipsoid~\cite{darnton2007torque}, and here we choose the sphere radius to be comparable to the cell's long semi-axis. This  assumption allows us to include hydrodynamic interactions analytically  with no need for a mesh and the use of boundary element or similar methods. We also restrict our attention to two flagella in order to focus on the simplest possible setup where hydrodynamic interactions can be measured; real \emph{E.~coli} features between two and five flagella~\cite{ping2010asymmetric}.

 	
	We simulate the cell and flagella dynamics for axial lengths $L=4,6,\SI{8}{\micro\metre}$ and base separation angles $\theta=\pi/8,\pi/4,3\pi/8,\pi/2$. 
 	An axial length $L=\SI{6.0}{\micro\metre}$ corresponds to a filament length of $\Lambda\approx\SI{6.9}{\micro\metre}$,   close to the average for \emph{E.~coli}~\cite{darnton2007torque}. The values $L=\SI{4}{\micro\metre}$ and $\SI{8}{\micro\metre}$ correspond to short and long flagella respectively.

\begin{figure}[t]
	\centering
	\begin{subfigure}{0.45\linewidth}
		\includegraphics[width=\textwidth]{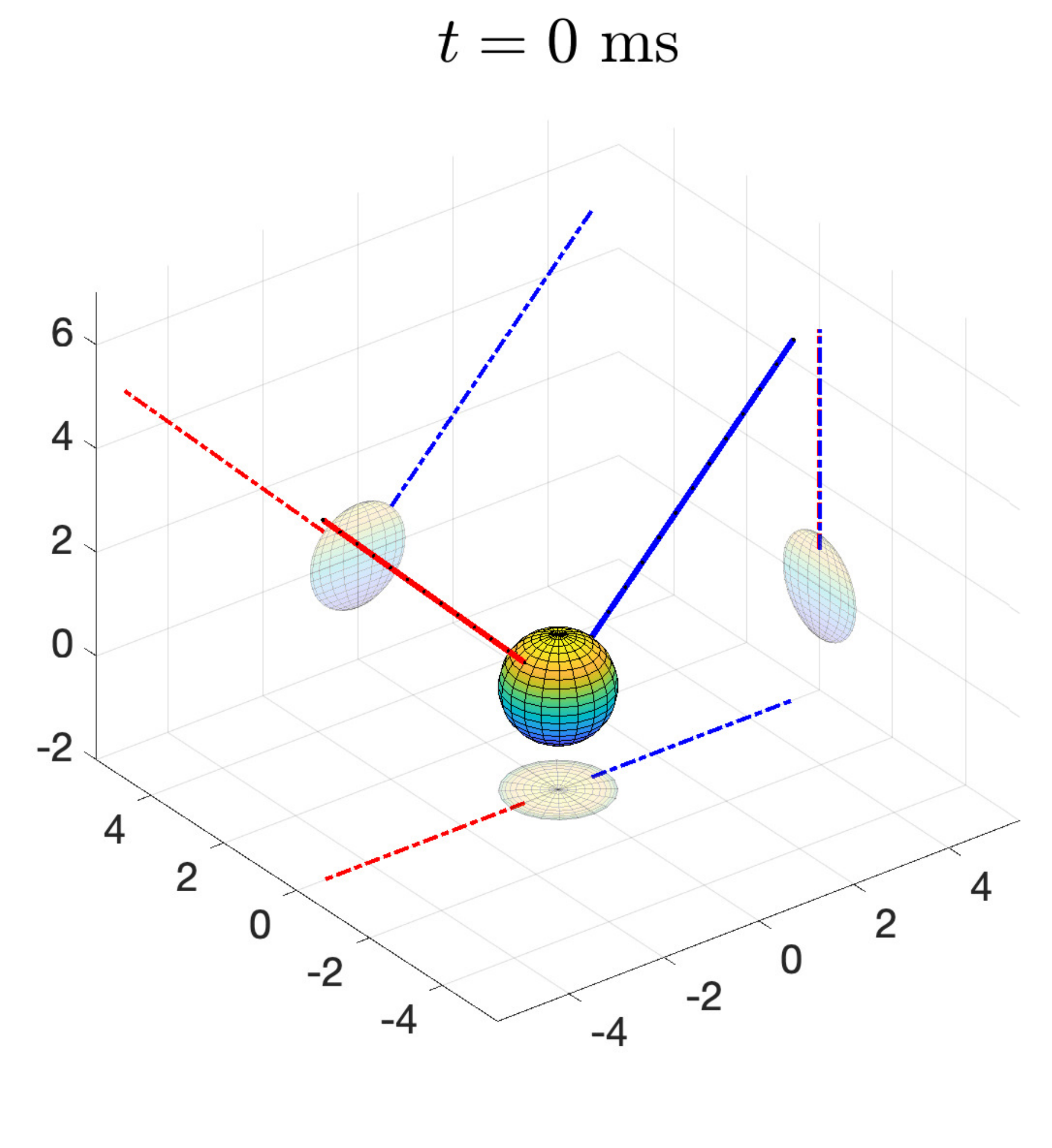}
	\end{subfigure}
	\hfill
	\begin{subfigure}{0.45\linewidth}
		\includegraphics[width=\textwidth]{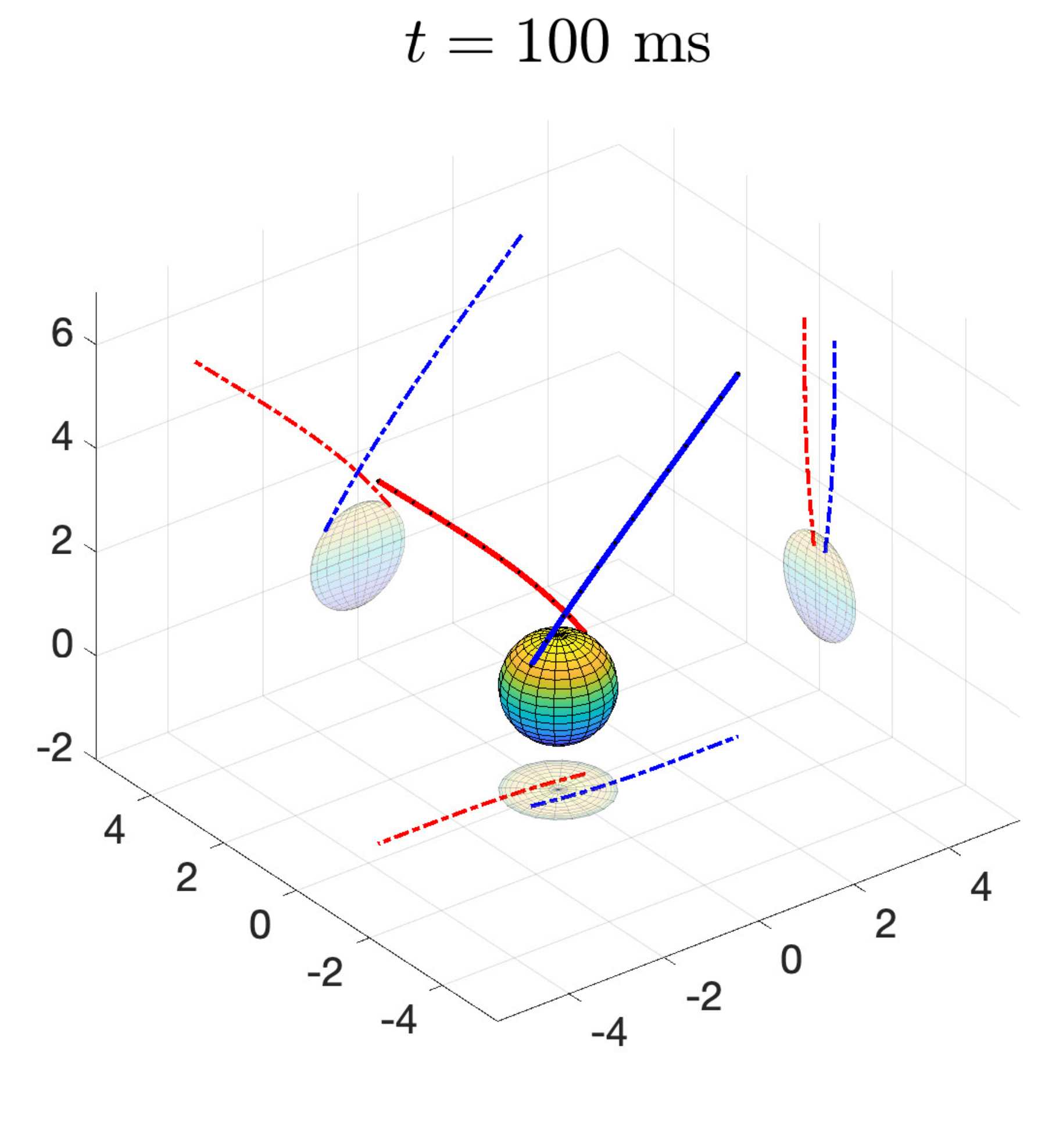}
	\end{subfigure}
	\hfill
	\begin{subfigure}{0.45\linewidth}
		\includegraphics[width=\textwidth]{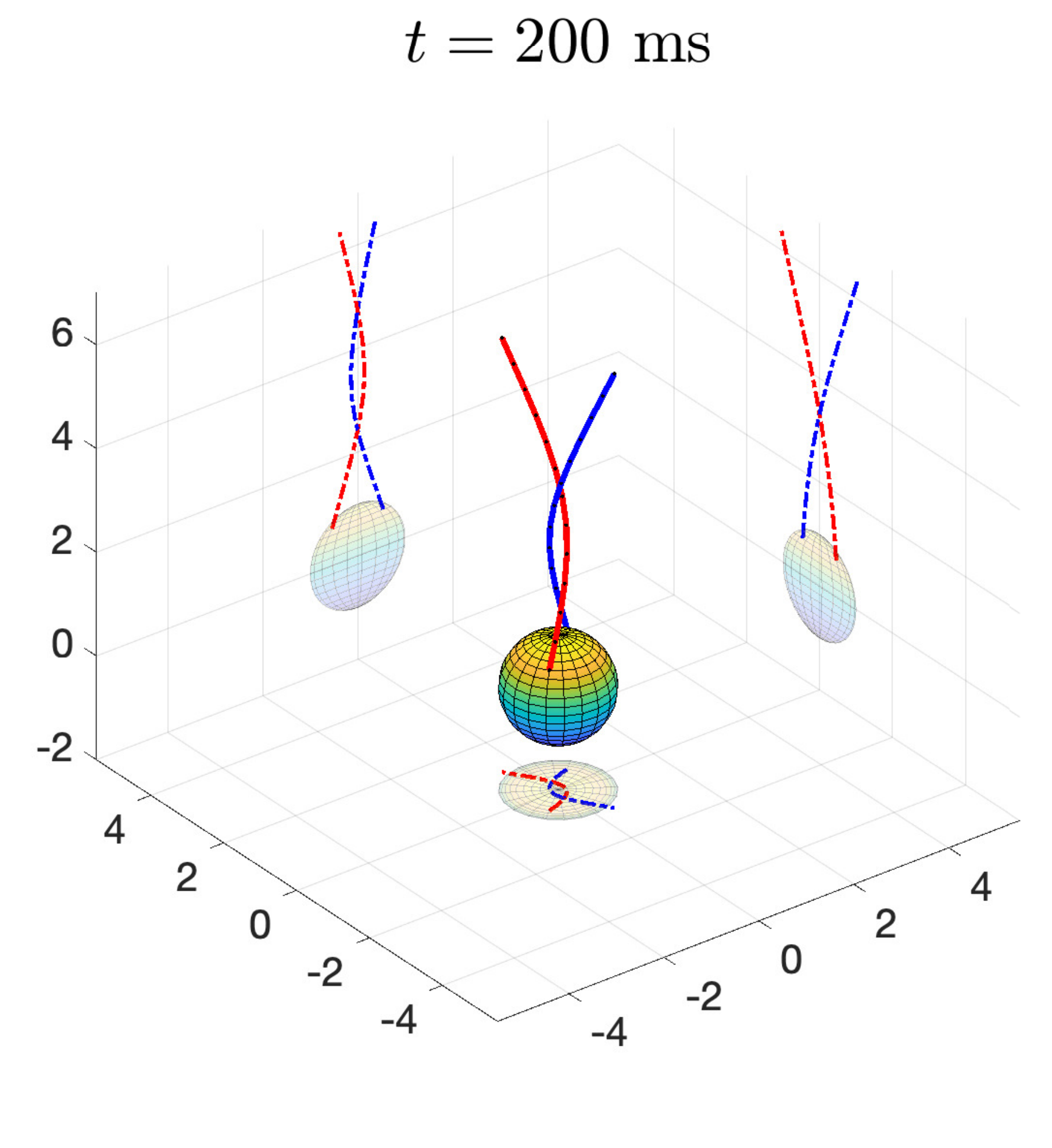}
	\end{subfigure}
	\hfill
	\begin{subfigure}{0.45\linewidth}
		\includegraphics[width=\textwidth]{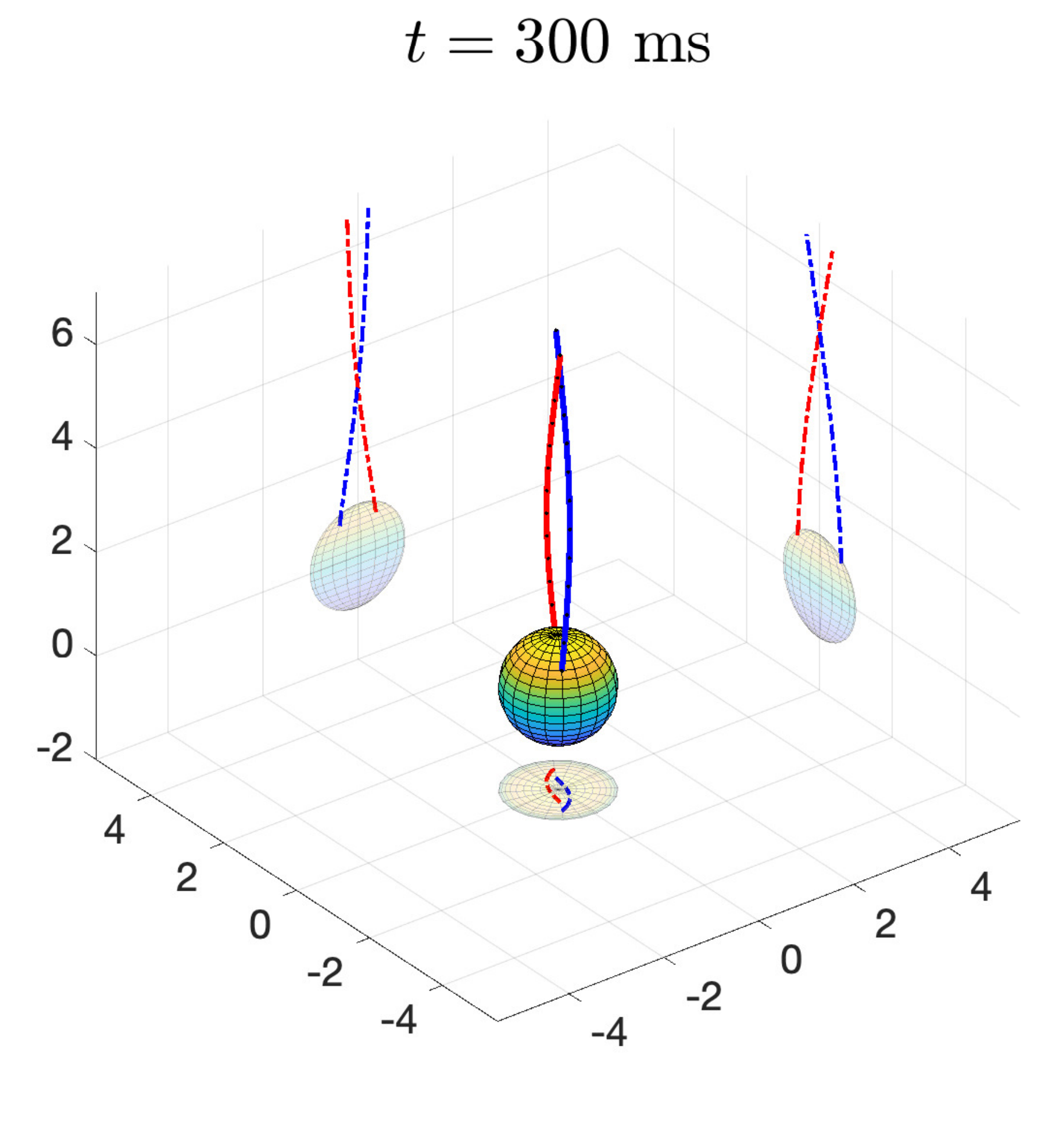}
	\end{subfigure}
	\caption{Snapshots of the bundling dynamics for $L=\SI{6}{\micro\metre}$ and $\theta=\pi/2$ viewed in the laboratory frame. The axes of the helical filaments are plotted in red and blue respectively with black dots representing hinges between segments. Shadows are drawn in dashed lines on three orthogonal planes to aid visualisation.}\label{bund:fig:snaps}
	\end{figure}

	First, in order to gain a general understanding of the bundling process we analyse the dynamics for $L=\SI{6}{\micro\metre}$ and $\theta=\pi/2$ in detail, before considering their variation for different values of $L$ and $\theta$. The filament axes are initialised as pointing straight radially outwards. In Fig.~\ref{bund:fig:snaps} we plot snapshots of the resulting bundling dynamics, as viewed in the laboratory frame. As the cell body rotates, the orientation of the axis segments remains nearly fixed in space, but due to cell body rotation they are being pulled at the base. After half a revolution, the filaments are sufficiently close for direct hydrodynamic interactions to have an effect. Over the course of another revolution of the cell body the axes of the filaments then slowly wrap around each other to form a bundle. The entire process takes about $\SI{0.3}{\second}$, a result which is consistent with experimental observations~\cite{berg2008coli,darnton2007torque}. 
	
	\begin{figure}[t]
		\centering
		\includegraphics[width=.8\textwidth]{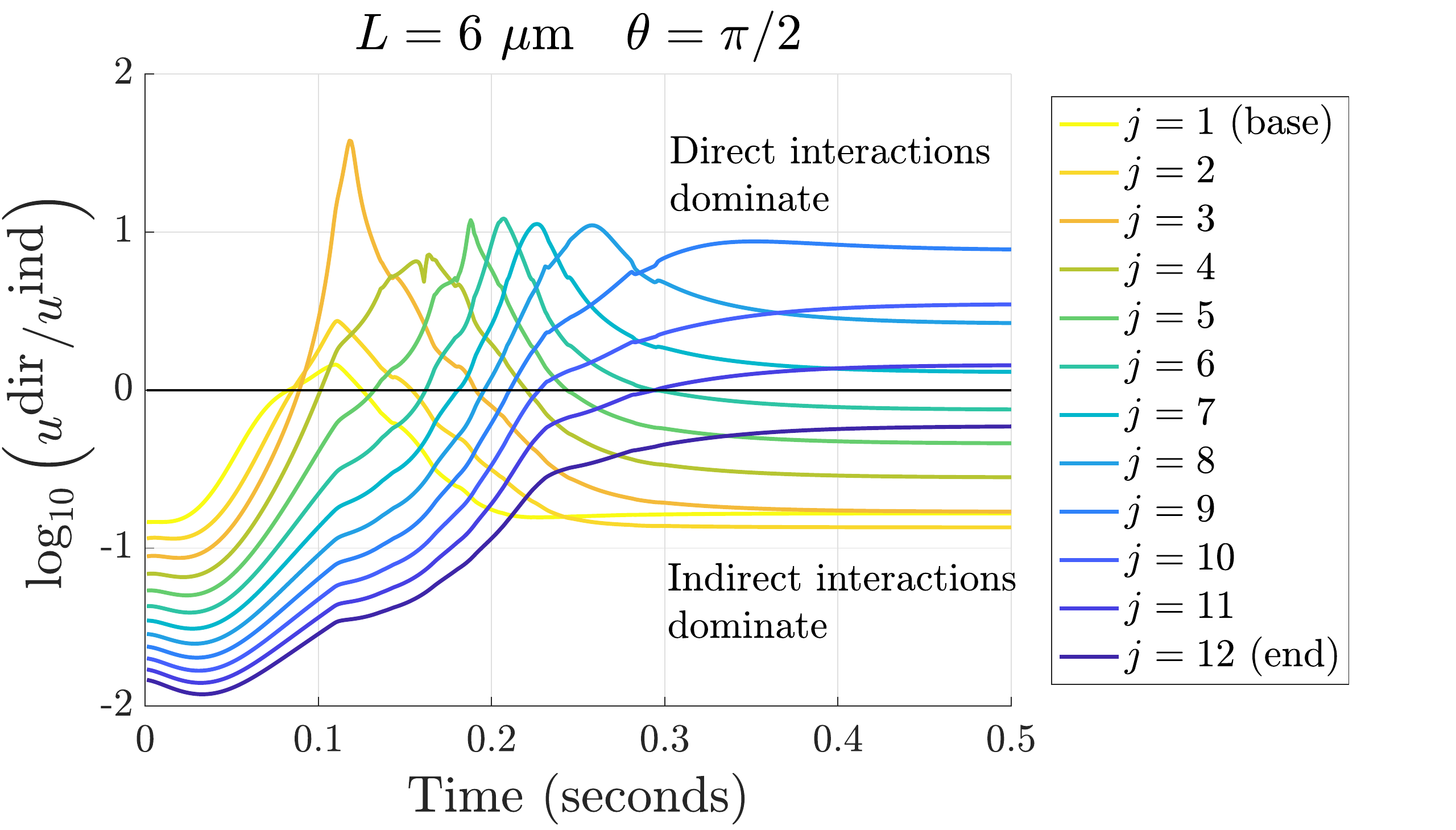}
		\caption{Ratio of direct and indirect interactions along the filament axis during the bundling process, obtained using the computational elastohydrodynamic model for filaments of axis length $L=\SI{6}{\micro\metre}$ and base separation angle $\theta=\pi/2$ with $N=12$ segments. Log-ratios for all segments from the base ($j=1$, yellow) to the end of the flagellum ($j=12$, blue) are shown.}\label{bund:fig:numresults1}
	\end{figure}

	We compute the ratio of direct to indirect advection at various points along the filament axis, and show the results in Fig.~\ref{bund:fig:numresults1}. Here each line, labelled by $j$, corresponds to one segment on the filament axis, with $j=1$ referring to the one linked to cell body, and $j=12$ referring to the free end of the flagellum. The quantities $u^{\text{dir}}$ and $u^{\text{ind}}$ are computed by projecting the corresponding velocity vector at $\bx_i^j$ into the plane perpendicular to $\bp_i^j$ and calculating the norm of that projection. In this analysis we combine force- and torque-induced advection into one quantity; a study  where forces and torques are computed separately yields a picture that is qualitatively similar to the one shown in Fig.~\ref{bund:fig:numresults1}.
	
	We then observe in Fig.~\ref{bund:fig:numresults1} that during the bundling process indirect interactions dominate everywhere along the filaments at early stages, consistent with the prediction from the minimal model. Moreover, the dominance is stronger at the outer ends, again consistent with our theoretical results. As the bundling process progresses, direct interactions become gradually more important. Interestingly, they peak in each case, first at the base and then with delay further along the filament. For $j\geq10$ the balance shift is monotonic, and along the entire filament the ratio plateaus once the bundle is formed. Surprisingly, the ratio is not monotonic along the filament in this final state. The dominance of direct effects is strongest for $j=9$, three quarters along the filament, while at the base and at the free end the balance is even shifted again in favour of indirect contributions. At the base this is due to the tether preventing a close approach. On the other hand, the ends of the filaments are the furthest away from the cell body centre and hence subject to the strongest indirect flows. The stiffness of the filaments is another contributing factor to this effect, since the absence of a bending moment at the free end requires the filaments to be straight there, as can also be seen in Fig.~\ref{bund:fig:snaps}.

	\begin{figure}[t]
		\centering
		\begin{subfigure}{0.24\linewidth}
			\includegraphics[width=\textwidth]{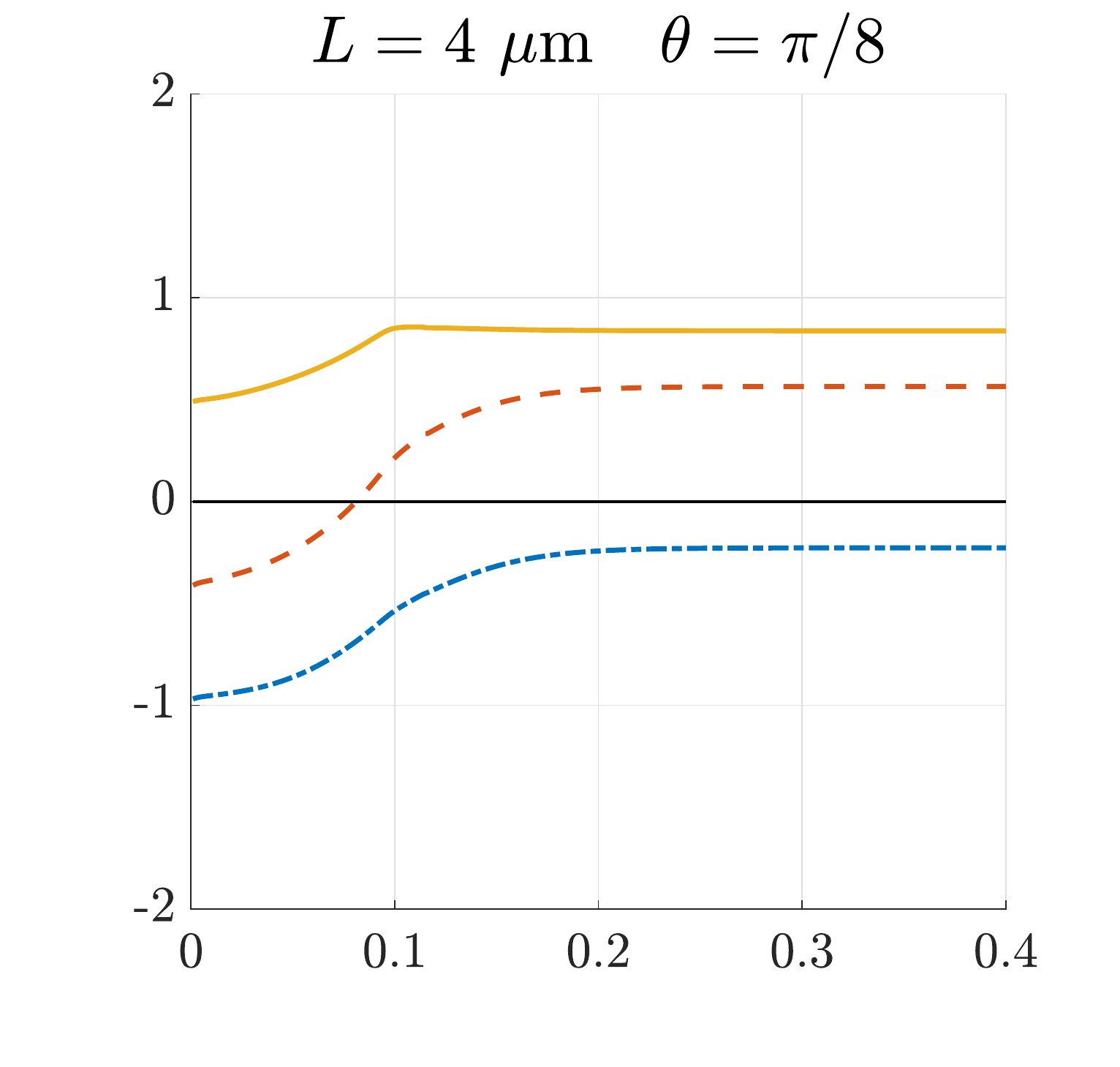}
		\end{subfigure}
		\hfill
		\begin{subfigure}{0.24\linewidth}
			\includegraphics[width=\textwidth]{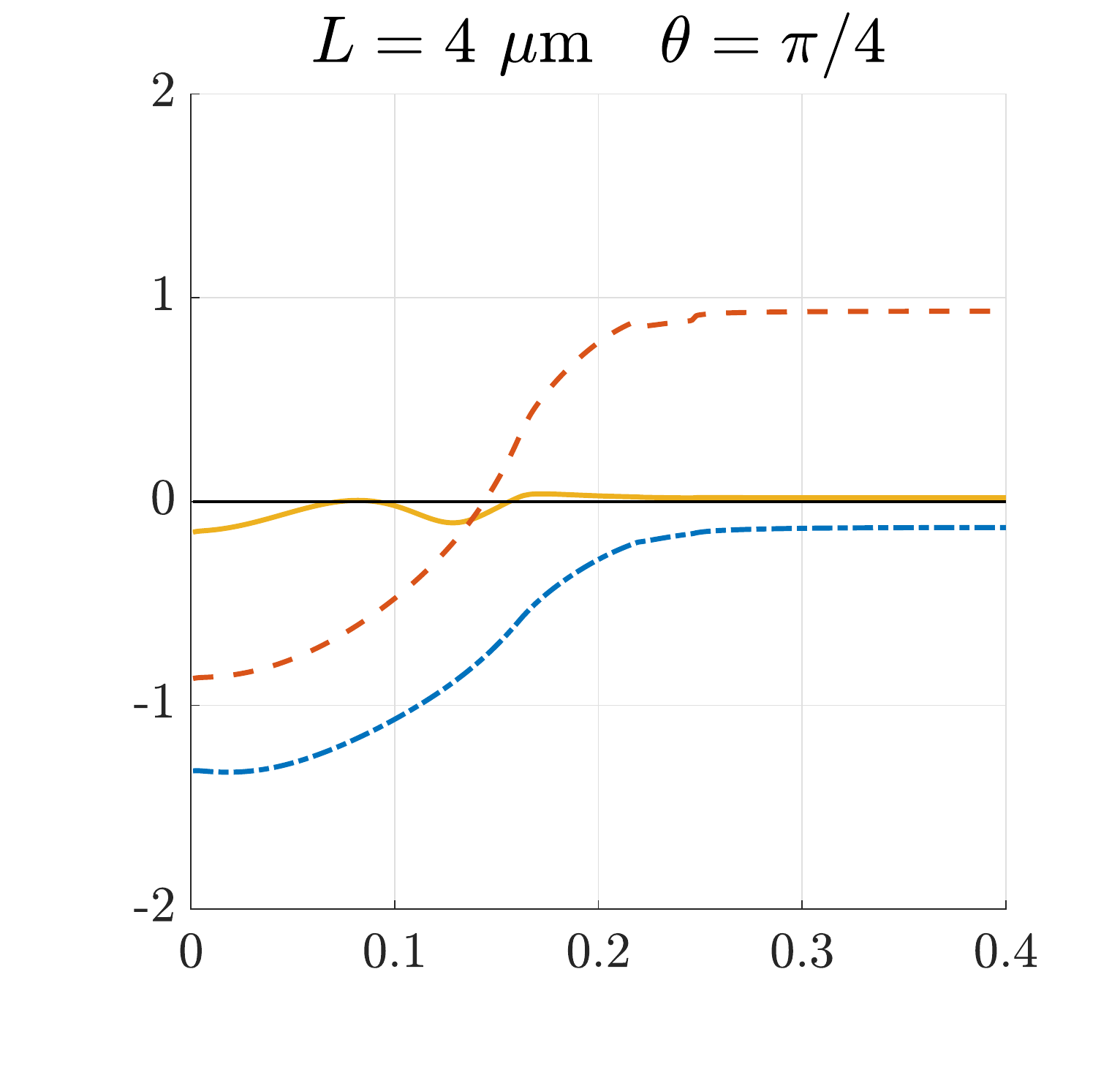}
		\end{subfigure}
		\hfill
		\begin{subfigure}{0.24\linewidth}
			\includegraphics[width=\textwidth]{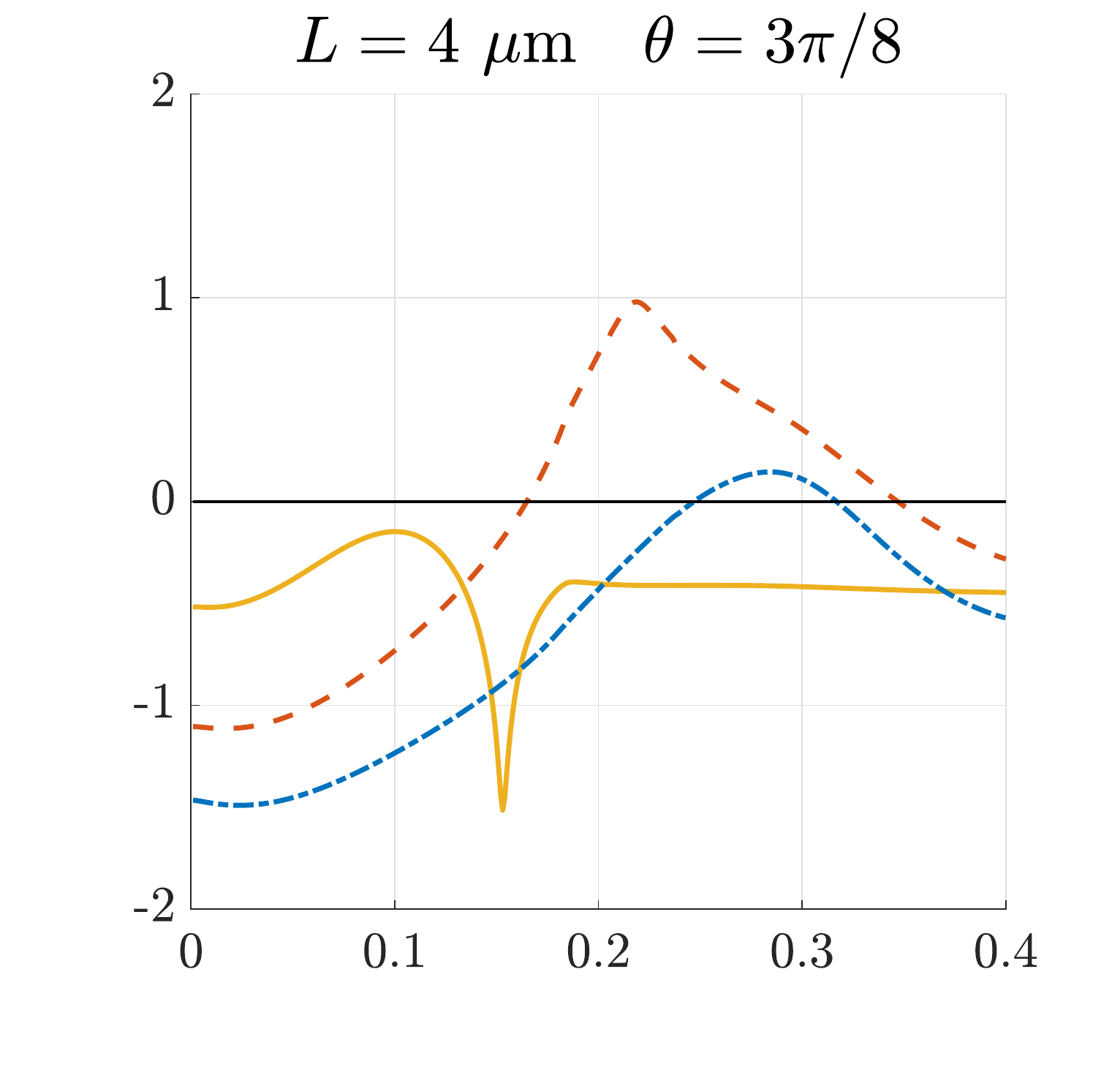}
		\end{subfigure}
		\hfill
		\begin{subfigure}{0.24\linewidth}
			\includegraphics[width=\textwidth]{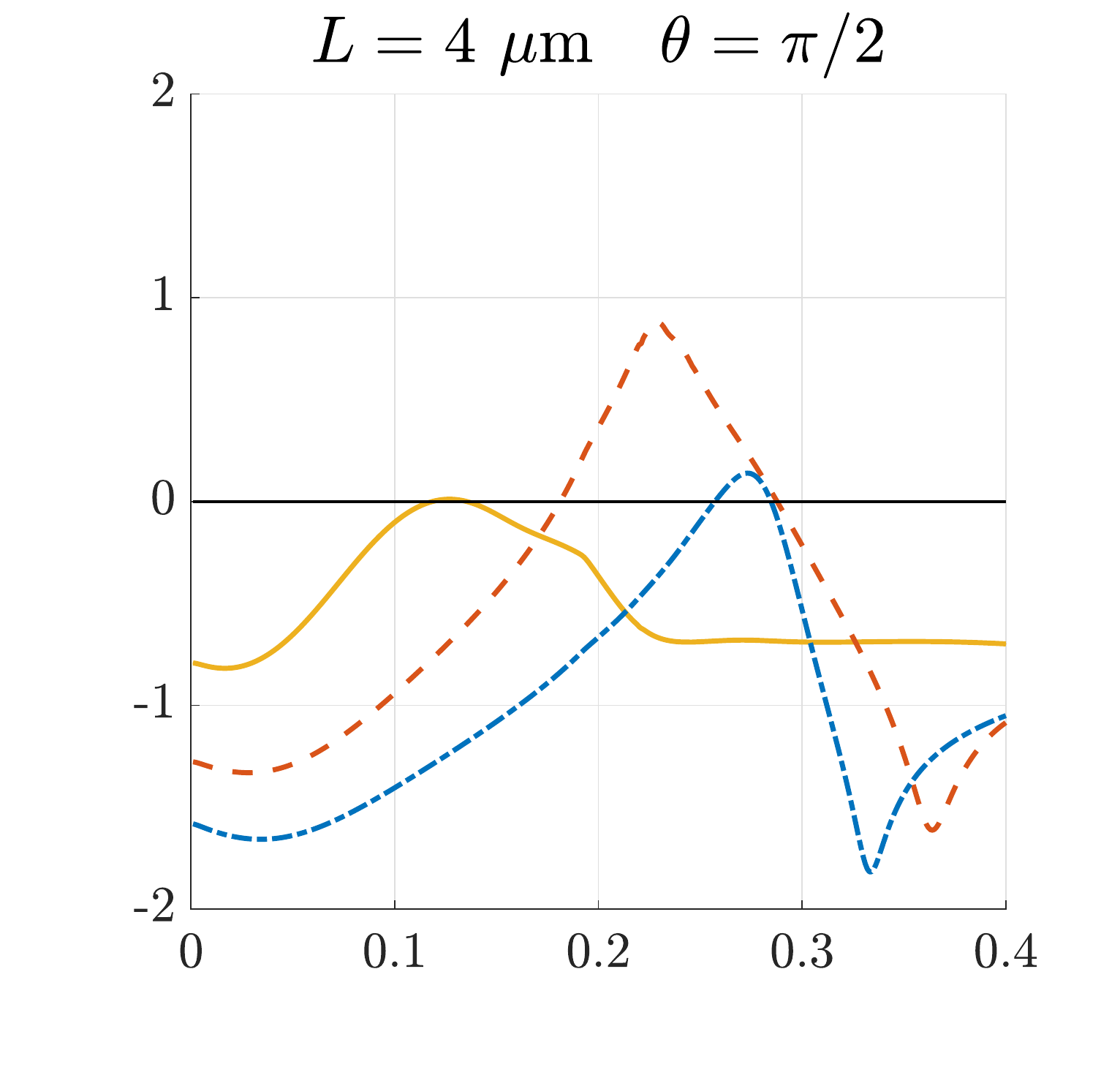}
		\end{subfigure}
		\hfill
		\begin{subfigure}{0.24\linewidth}
			\includegraphics[width=\textwidth]{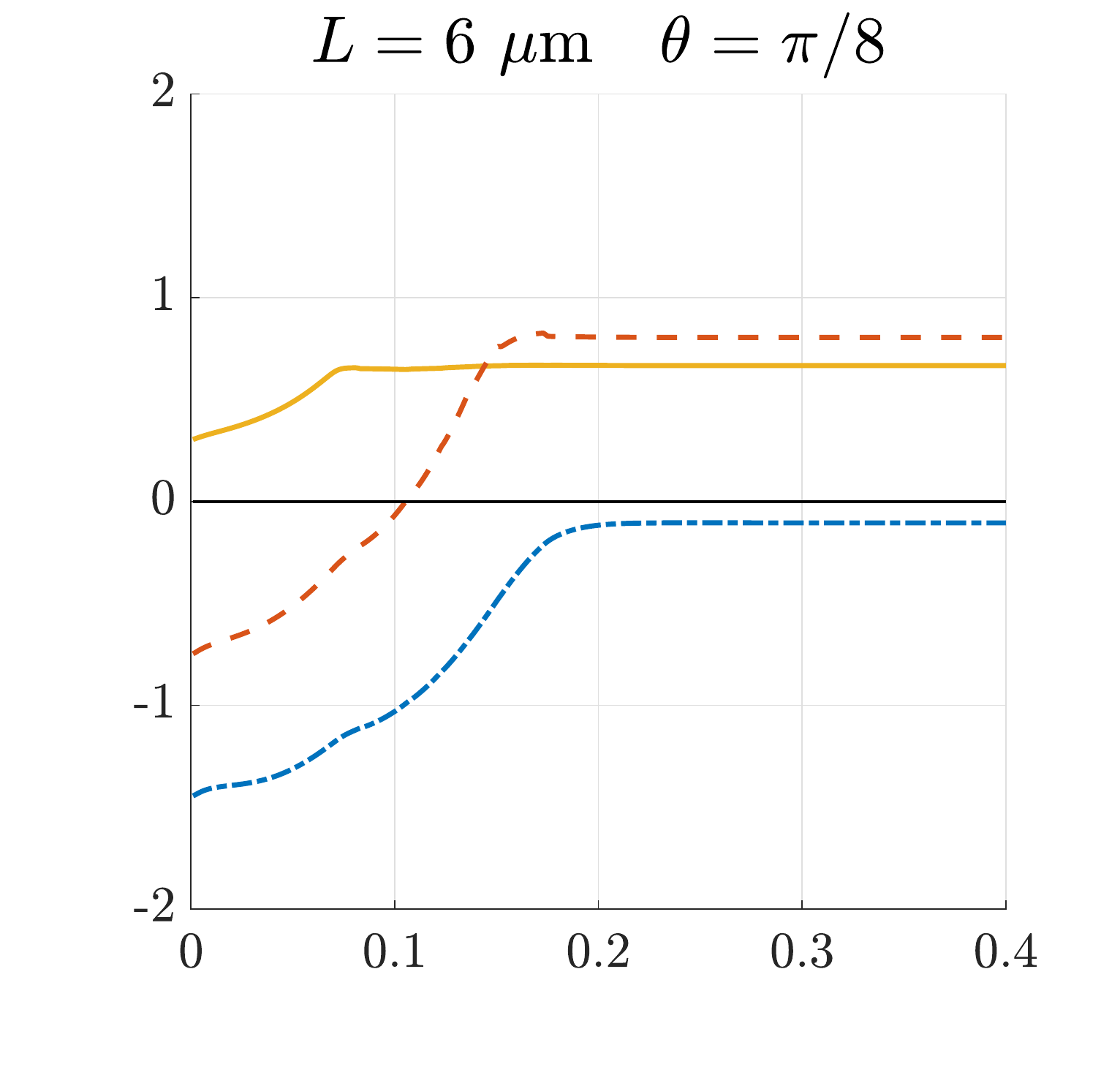}
		\end{subfigure}
		\hfill
		\begin{subfigure}{0.24\linewidth}
			\includegraphics[width=\textwidth]{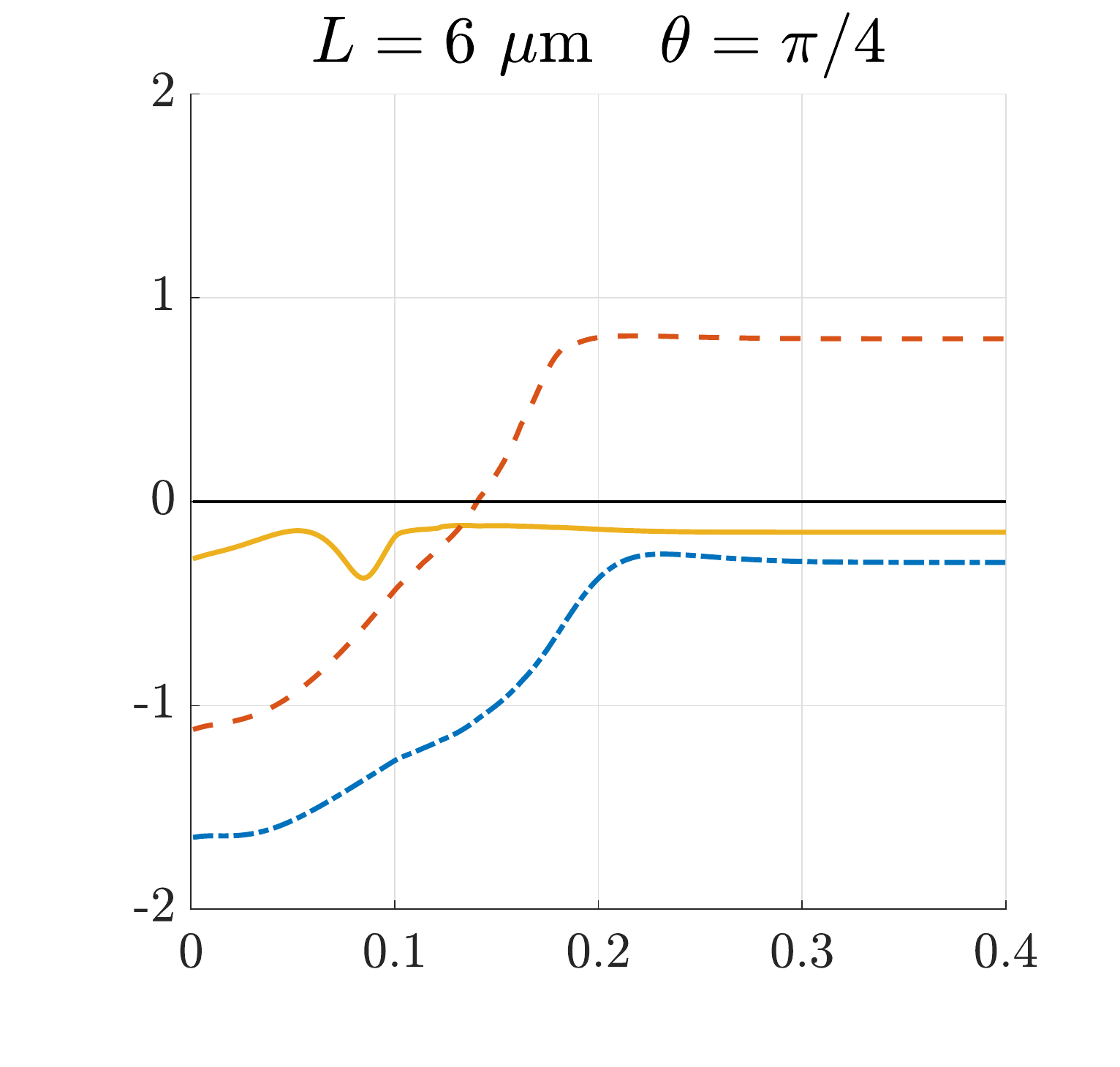}
		\end{subfigure}
		\hfill
		\begin{subfigure}{0.24\linewidth}
			\includegraphics[width=\textwidth]{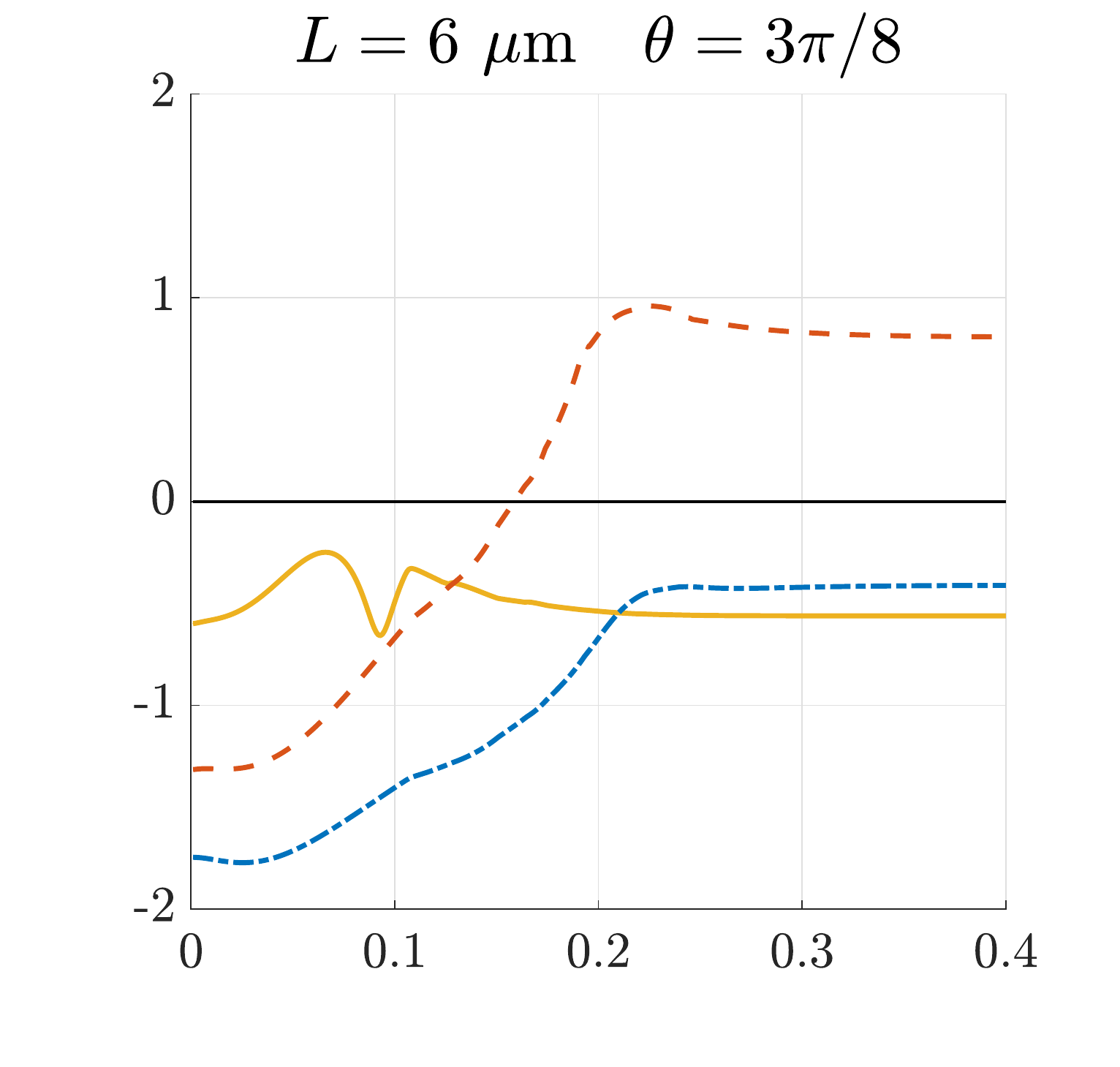}
		\end{subfigure}
		\hfill
		\begin{subfigure}{0.24\linewidth}
			\includegraphics[width=\textwidth]{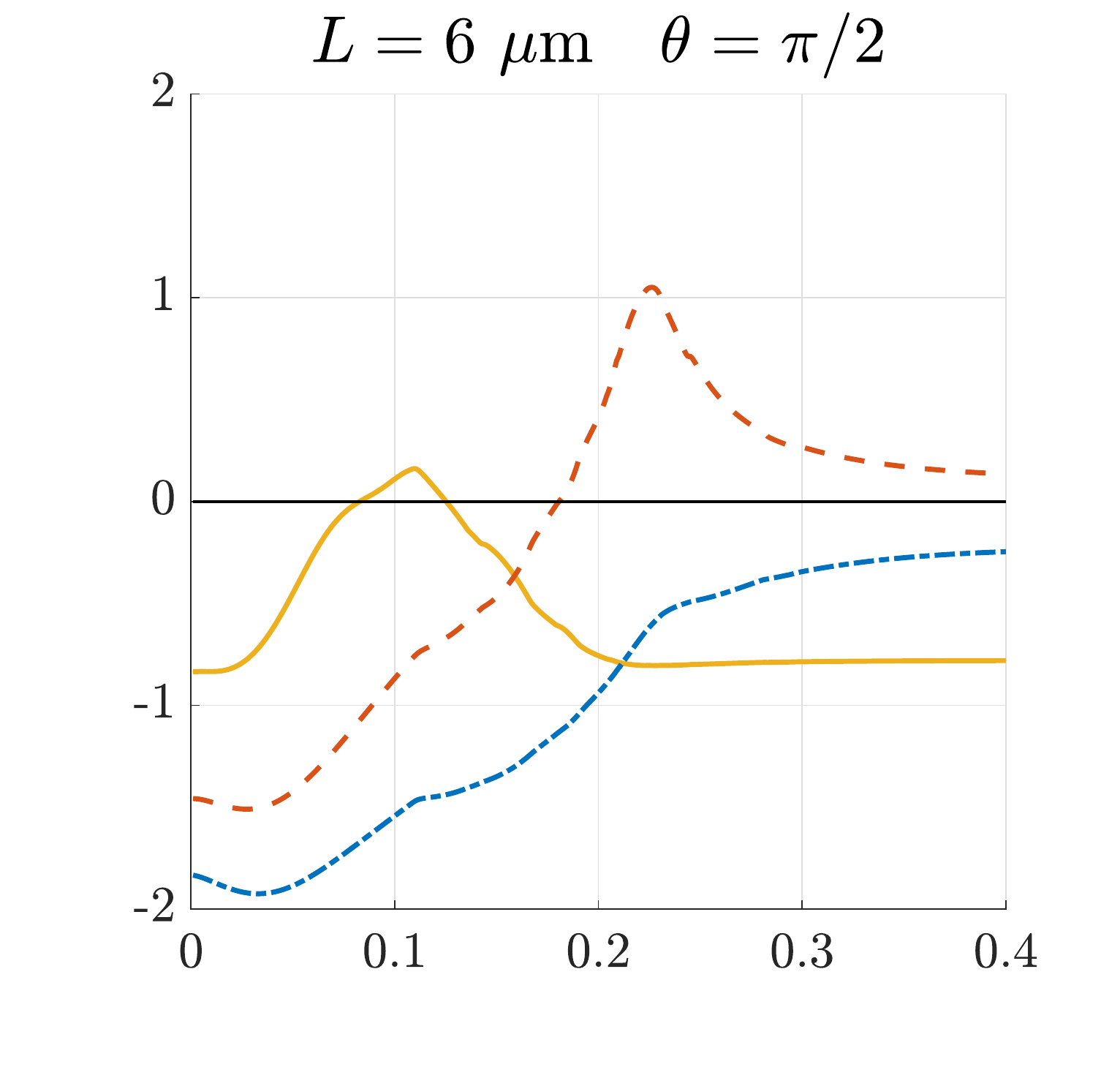}
		\end{subfigure}
		\hfill
		\begin{subfigure}{0.24\linewidth}
			\includegraphics[width=\textwidth]{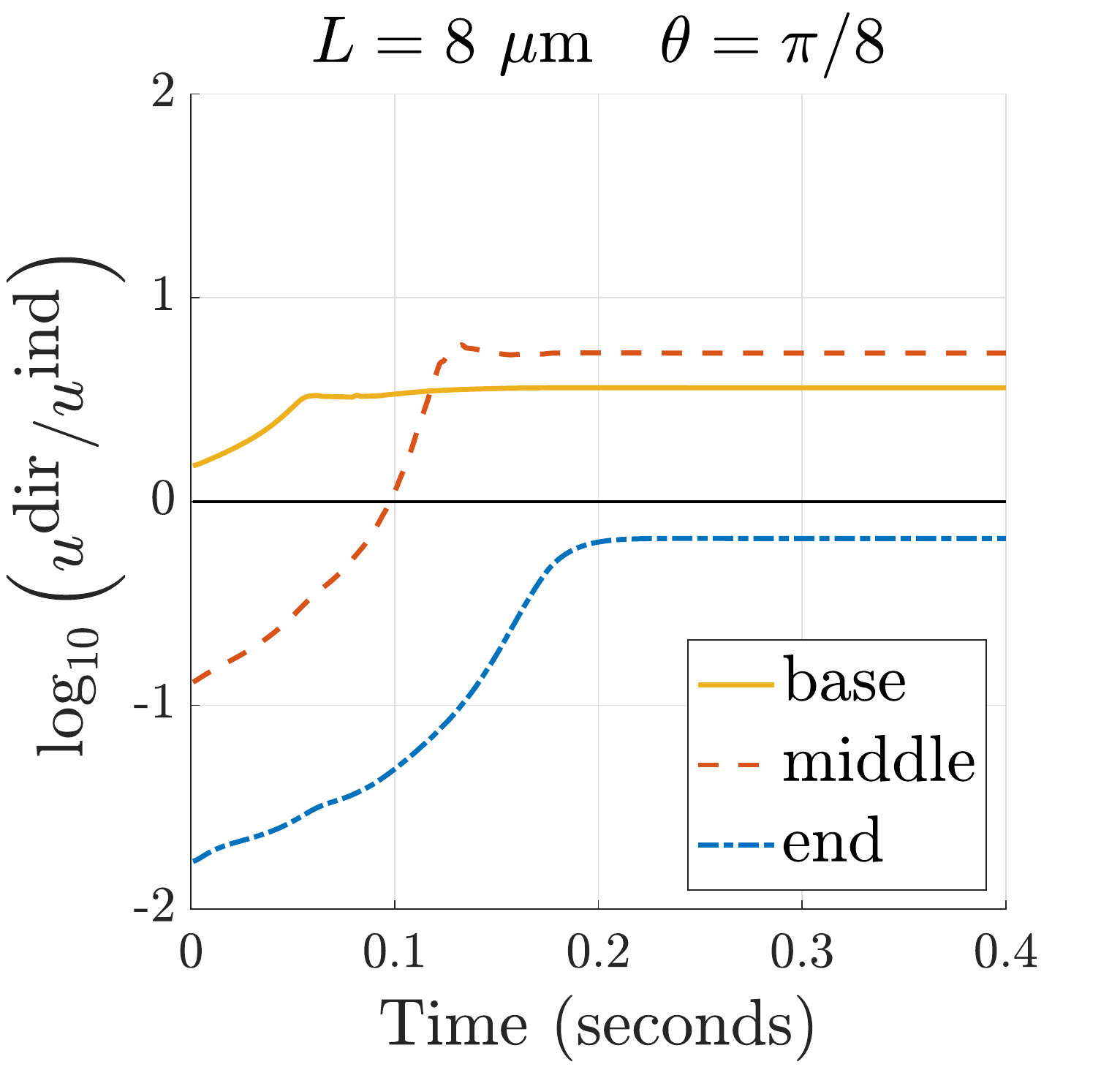}
		\end{subfigure}
		\hfill
		\begin{subfigure}{0.24\linewidth}
			\includegraphics[width=\textwidth]{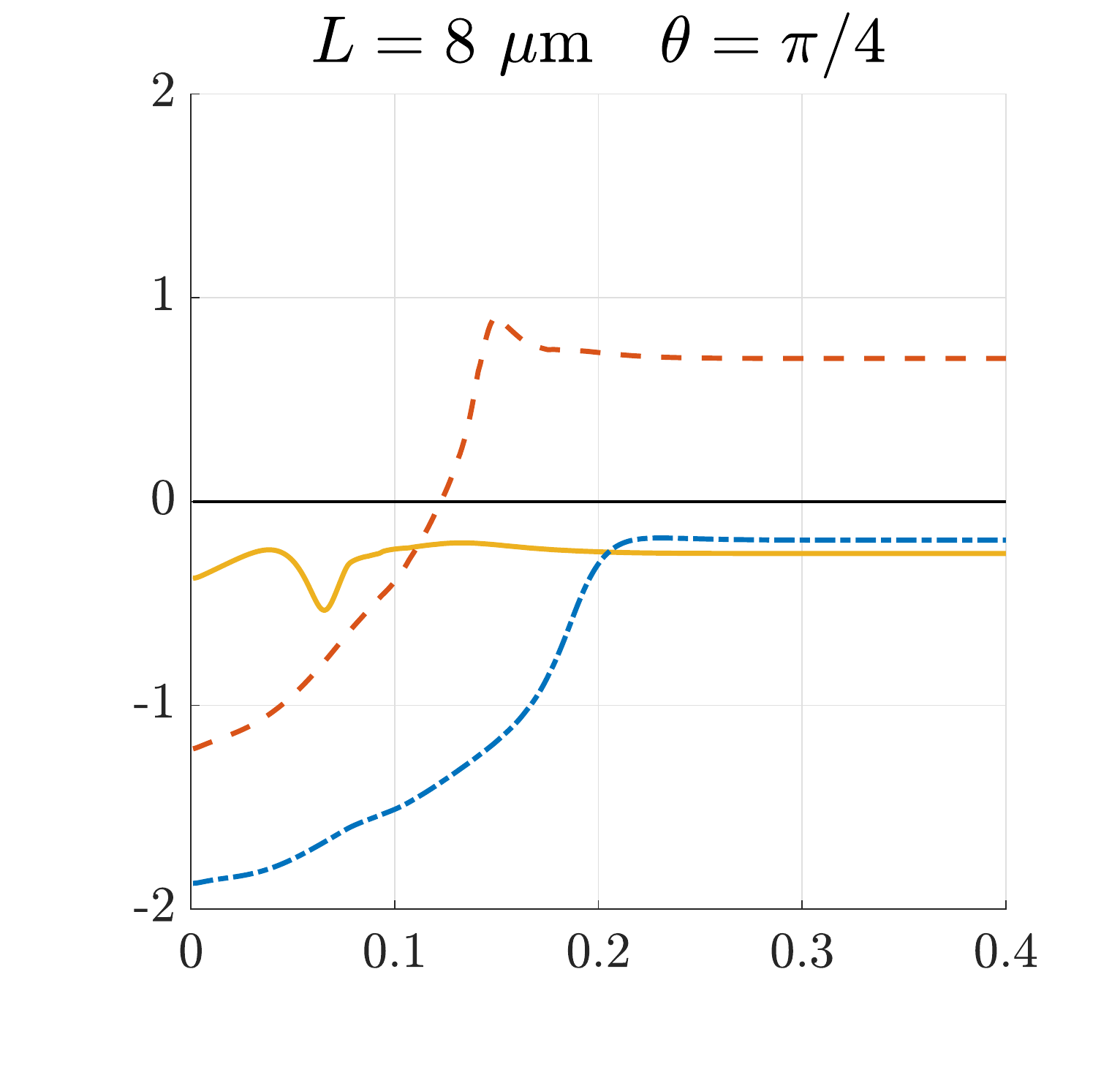}
		\end{subfigure}
		\hfill
		\begin{subfigure}{0.24\linewidth}
			\includegraphics[width=\textwidth]{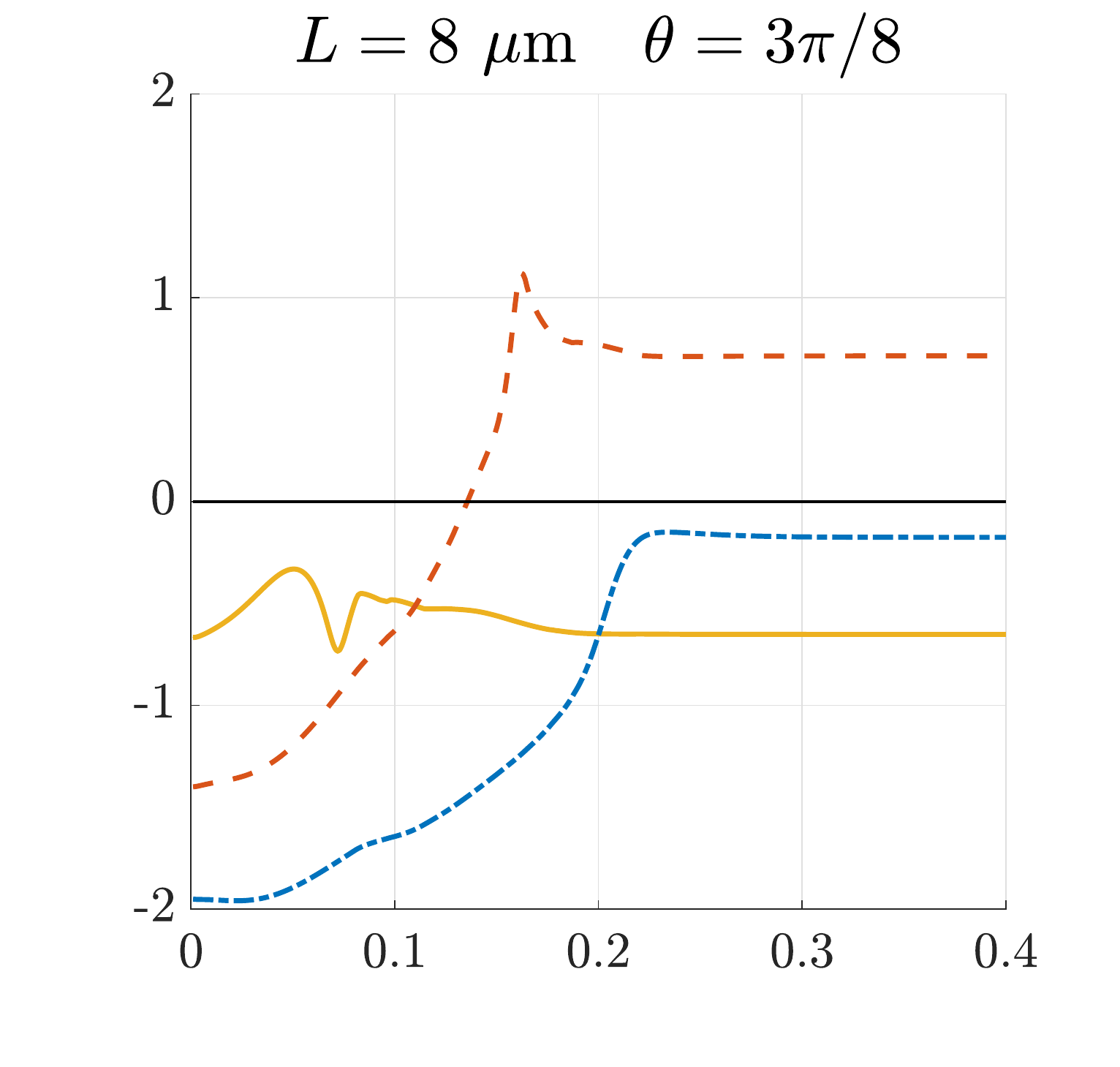}
		\end{subfigure}
		\hfill
		\begin{subfigure}{0.24\linewidth}
			\includegraphics[width=\textwidth]{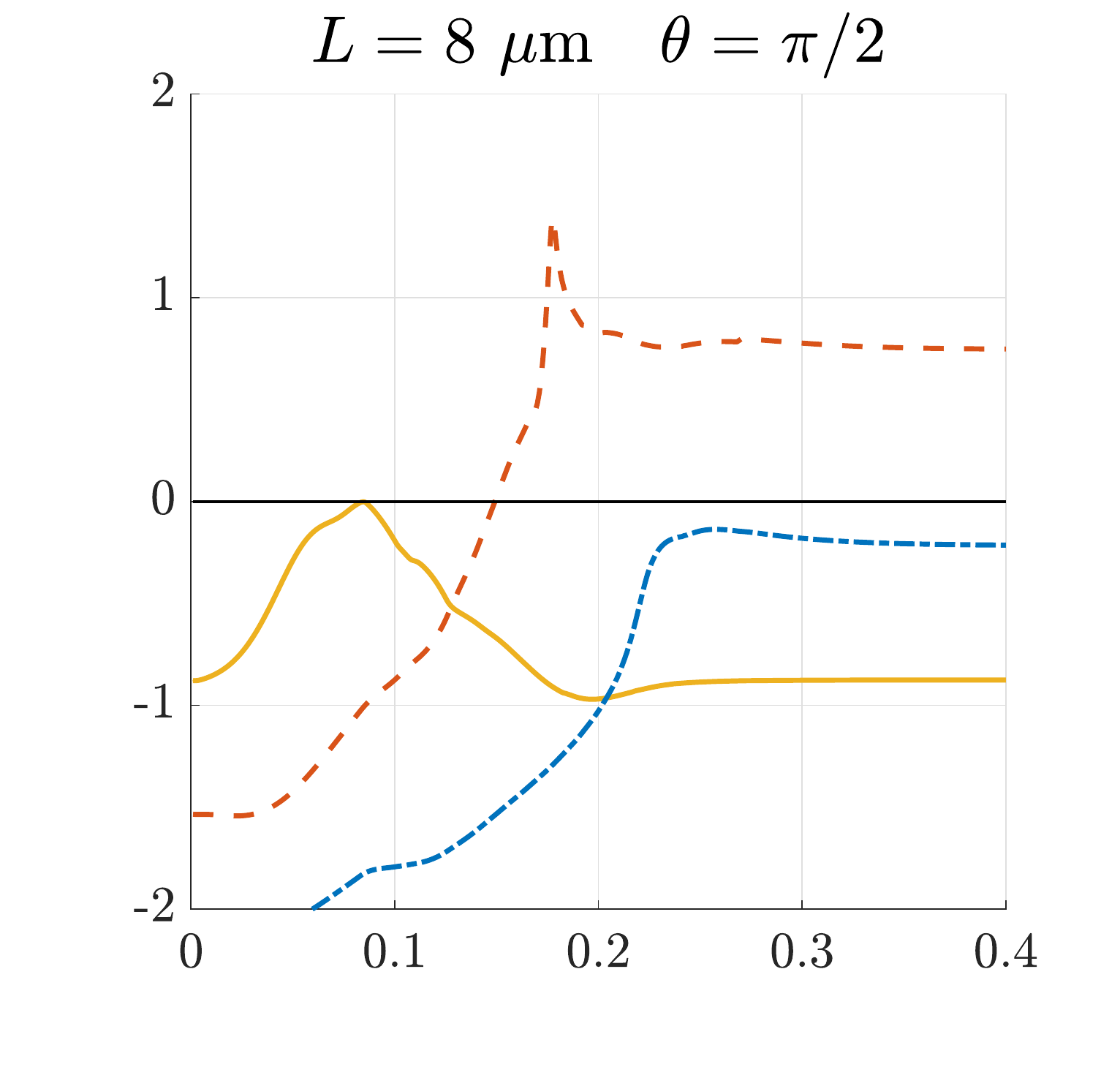}
		\end{subfigure} 
		\caption{Computed  bundling dynamics for filament axial lengths   $L=4$, $6$ and $\SI{8}{\micro\metre}$ (with $N=8$, $12$, $16$ respectively) and $\theta=\pi/8$, $\pi/4$, $3\pi/8$ and $\pi/2$. Lines are drawn for three representative values of $j$ at the base of the flagellar filament (yellow), middle (red, dashed) and end (blue, dash-dotted), which are $j=\{1,5,8\}$, $\{1,7,12\}$ and $\{1,9,16\}$ for $L=4$, $6$, $\SI{8}{\micro\metre}$ respectively.}\label{bund:fig:all}
	\end{figure}

	Having analysed one case in detail, we now ask how the bundling dynamics depend on the geometric parameters. To this end we illustrate in Fig.~\ref{bund:fig:all} results for different values of the filament axial length $L$ and the angle between the motors $\theta$. For clarity we only plot three lines in each case, representing the base, middle and end of the flagellum respectively (the specific values of the element $j$ plotted are listed in the caption of Fig.~\ref{bund:fig:all}). 	
	
	At close separation, $\theta=\pi/8$, the dynamics are very similar for different filament lengths. In each case the shift from indirect to direct interactions is nearly monotonic, and occurs first at the base, then in the centre and finally at the end. In the case of long flagella however, we observe that in the final state direct interactions are strongest close to the centre points of the filament. This effect is significantly more pronounced when the filaments are more widely separated at their base. This is due to the stiffness of the filaments, which makes it harder to align their free ends if they are bundled at an angle.
	
	For short filaments, $L=\SI{4}{\micro\metre}$, we note that a bundle fails to form if $\theta\geq3\pi/8$. The filament axes touch after half a revolution of the cell body but instead of wrapping they slide past each other until they reach an aligned steady state. During this sliding process the separation between them increases, which explains the non-monotonic behaviour of the advection ratio.
	
	{
	In addition to the results shown above, we also experimented with filaments of different lengths and with different strengths of the singularity distributions,   capturing the effect that varying the flagellar geometry would have according to Eqs.~\eqref{bund:eq:Fi} and \eqref{bund:eq:Gi}. 	
	For small perturbations from the symmetric case, the dynamics at early stages (i.e.~during the bundle formation) are seen to be qualitatively the same as in the  case of two symmetric filaments detailed above. After the formation of a bundle, two asymmetric filaments can be   seen to tilt laterally, while remaining stably interlocked as a bundle. This is due to unbalanced parallel torques, which generate a net translation (as is the case next to a planar interface as well). 
		
	This observation puts further emphasis on the limitation of any singularity-type model at late times. When the filaments are widely separated, i.e.~during the early stages of bundling, the dynamics are robust to perturbations in the filaments characteristics, and so our conclusion that indirect interactions are dominant remains true. Once the bundle is formed, the assumption of a constant force and torque singularity distribution along the filament axis  breaks down and, instead, short-ranged elastohydrodynamic interactions determine the force distribution dynamically. 	}

 	\section{Discussion}\label{bund:sec:discussion}
	
	In this paper we defined the notion of direct and indirect advection of bacterial flagellar filaments, and used it to classify the fundamental flows contributing to the bundling process. In addition we derived a computational elastohydrodynamic model that corroborates our predictions and leads to new ones.
	
	First, we saw that the generation of thrust by flagella is sufficient to initiate bundling through hydrodynamic interactions. This follows from the fact that both the direct and indirect interactions, computed in Eq.~\eqref{bund:eq:ulat_A} and Eq.~\eqref{bund:eq:ulat_P} respectively, lead to non-zero advection towards a bundle in the entire geometric parameter range. Secondly, for both thrust- and rotation-induced flows indirect advection always dominates for long filaments and at wide separation, i.e.~during the early stages of the bundling process. The shape of the boundary between direct- and indirect-dominated regions suggests a `wake' that a flagellar filament is generating behind the cell body with a width on the order of the cell body diameter. In other words, indirect interactions dominate unless the filaments are in each other's wake. Thirdly, in terms of the dynamic bundling process, the relative contribution of direct and indirect effects was seen to depend on the length of the flagellum, with active contributions being more important the shorter the flagellar filaments are. 
	
	Fourthly, the rotation of the flagella only generates flows in the azimuthal direction and thus does not facilitate bundling purely through hydrodynamic effects. However, as evidenced by our computational model, they do play a role in conjunction with other physical constraints, such as tethering. This was already known for direct hydrodynamic interactions~\cite{kim2003macroscopic,man2016hydrodynamic}, and here we demonstrate that it holds true for indirectly generated flows as well. Finally, from the computational model we see  that the shift in balance between direct and indirect effects is non-monotonic during the bundling process, with a peak in direct dominance, and furthermore that different sections of the flagella are affected by these changes to different extents.
	
	Naturally, by reducing the system to two geometric parameters we made a number of simplifying assumptions. In reality, cells are slightly denser than water (for \emph{E.~coli} {the density $\varrho\approx1.08~\si{\gram\per\milli\litre}$~\cite{kubitschek1983buoyant} while for water $\varrho\approx1.00~\si{\gram\per\milli\litre}$}), and their cell body is normally better approximated by a prolate spheroid and they feature more than two flagella~\cite{berg2008coli}. Furthermore, we restricted our attention to two flagella, while \emph{E.~coli} features between two and five~\cite{ping2010asymmetric}. A particularly interesting  direction to explore with this framework would be the dynamics of more than two flagella, since  in that case   much of the symmetry of this problem is lost, which could lead to more complex dynamics. 	In addition, our model might lead to a better understanding of the unbundling process, which triggers the `tumble' gait of multi-flagellated bacteria. When a peritrichous bacterium reverses the direction of one flagellar motor to clockwise rotation, the torque exerted by that flagellum is reversed. However, the total torque, and hence the strength of the indirect advection, is weakened but not cancelled due to continuing counter-clockwise rotation of the remaining flagella, thus acting to keep the bundle together at the end. These indirect effects may therefore provide the physical mechanism for the apparent  `pulled out' motion of the clockwise rotating flagellum observed at the base of the flagellar bundle~\cite{darnton2007torque}.

\section*{Acknowledgements}
This project has received funding from the European Research Council (ERC) under the European Union's Horizon 2020 research and innovation programme  (grant agreement 682754 to EL).

	\bibliographystyle{ieeetr}
	\bibliography{bundling}
	
\end{document}